\input harvmac
\input amssym.def
\input amssym.tex
\input epsf.tex

\noblackbox

\def\AdS{$AdS$}
\def\CFT{$CFT$}
\def\bk{{\bf N}}
\def\bkb{{\bf\overline N}}
\def\dP{$dP_3$}

\def\BC{{\Bbb C}}
\def\BF{{\Bbb F}}
\def\BP{{\Bbb P}}
\def\BR{{\Bbb R}}
\def\BZ{{\Bbb Z}}
\def\CB{{\cal B}}
\def\CC{{\cal C}}

\def\CL{{\cal L}}
\def\CM{{\cal M}}
\def\CN{{\cal N}}
\def\CO{{\cal O}}
\def\CR{{\cal R}}
\def\RH{{\rm H}}

\def\urlfont{\hyphenpenalty=10000
\hyphenchar\tentt='057 \tt }

\newbox\tmpbox\setbox\tmpbox\hbox{\abstractfont DUKE-CGTP-01-11}
\newbox\tmpbox\setbox\tmpbox\hbox{\abstractfont PUPT-2006}
\Title{\vbox{\baselineskip12pt\hbox{\hss hep-th/0109053}
\hbox{DUKE-CGTP-01-11}\hbox{PUPT-2006}}}
{\vbox{
\centerline{Toric Duality Is Seiberg Duality}}}
\smallskip
\centerline{Chris E. Beasley\foot{cbeasley@princeton.edu}}
\smallskip
\centerline{\it{Joseph Henry Laboratories, Princeton University, Princeton, New Jersey 08544}}
\medskip
\centerline{and}
\medskip
\centerline{M. Ronen Plesser\foot{plesser@cgtp.duke.edu}}
\smallskip
\centerline{\it{Center for Geometry and Theoretical Physics, Box
90318, Duke University, Durham, NC \ 27708-0318}}
\bigskip\bigskip

We study four $\CN=1$ $SU(N)^6$ gauge theories, with bi-fundamental
chiral matter and a superpotential.  In the infrared, these gauge
theories all realize the low-energy world-volume description of $N$
coincident D3-branes transverse to the complex cone over a del
Pezzo surface \dP\ which is the blowup of $\BP^2$ at three generic points. 
Therefore, the four gauge theories are expected to fall into the 
same universality class---an example of a phenomenon that has been 
termed ``toric duality.''  However, little independent evidence has 
been given that such theories are infrared-equivalent.

In fact, we show that the four gauge theories are related by the
$\CN=1$ duality of Seiberg, vindicating this expectation.  We also
study holographic aspects of these gauge theories.  In particular we
relate the spectrum of chiral operators in the gauge theories to
wrapped D3-brane states in the \AdS\ dual description.  We finally
demonstrate that the other known examples of toric duality are related
by $\CN=1$ duality, a fact which we conjecture holds generally.
\bigskip

\Date{September 2001}

\lref\ceb{C.E. Beasley, ``Superconformal Theories from Branes at
Singularities,'' Duke University senior honors thesis, 1999.}

\lref\agm{
P.~S. Aspinwall, B.~R. Greene, and D.~R. Morrison, ``Calabi-Yau moduli
space, mirror manifolds, and spacetime topology change in string
theory,'' Nucl. Phys. B {\bf 416} (1994) 414--480, {\urlfont
hep-th/9309097}.}

\lref\afgj{
D. Anselmi, D.~Z. Freedman, M.~T. Grisaru, and A.~A. Johansen,
``Nonperturbative Formulas for Central Functions in Supersymmetric
Gauge Theories,''  Nucl. Phys. B {\bf 526} (1998) 543--571, {\urlfont
hep-th/9708042}.}

\lref\bg{
C.~P. Boyer and K. Galicki, ``On Sasakian-Einstein Geometry,''
Int. J. Math. {\bf 11} (2000) 873--909, {\urlfont math.DG/9811098}.}

\lref\bglp{
C. Beasley, B.~R. Greene, C.~I. Lazaroiu, and M.~R.
Plesser,
``D3-branes on partial resolutions of abelian quotient
singularities in
Calabi-Yau threefolds,'' Nucl. Phys. B {\bf 566} (2000)
599--640,
{\urlfont hep-th/9907186}.}

\lref\bh{
A. Bergman and C. Herzog, ``The Volume of some Non-spherical Horizons
and the \AdS /CFT Correspondence,'' {\urlfont hep-th/0108020}.}

\lref\bt{
R. Bott and L. Tu, {\it Differential Forms in Algebraic Topology},
Springer, 1982.}

\lref\dgm{
M.~R. Douglas, B.~R. Greene, and D.~R. Morrison, 
``Orbifold resolution by D-branes,'' Nucl. Phys. B
{\bf 506} (1997) 84--106, 
{\urlfont hep-th/9704151}.}

\lref\dm{
M.~R. Douglas and G. Moore, 
``D-branes, quivers, and ALE instantons,'' {\urlfont
hep-th/9603167}.}

\lref\fhhtd{
B.~Feng, A.~Hanany, and Y.~He,
``D-brane Gauge Theories from Toric Singularities and Toric Duality,''
Nucl.\ Phys.\ B {\bf 595} (2001) 165--200, {\urlfont hep-th/0003085}.}

\lref\fhhps{
B.~Feng, A.~Hanany, and Y.~He, ``Phase Structure of D-brane Gauge
Theories and Toric Duality,'' J. High Energy Phys. {\bf 08} (2001) 040,
{\urlfont hep-th/0104259}.}

\lref\fulton{
W. Fulton, {\it Introduction to Toric Varieties}, Princeton Univ. Press, 1993.}

\lref\greene{
B.~R. Greene, ``D-brane topology changing
transitions,'' 
Nucl. Phys. B {\bf 525} (1998) 284--296, {\urlfont
hep-th/9711124}.}

\lref\g{
S.~S. Gubser, ``Einstein Manifolds and Conformal Field Theories,''
Phys. Rev. D {\bf 59} (1999) 025006, {\urlfont hep-th/9807164}.}

\lref\sem{
N. Seiberg, ``Electric-magnetic duality in supersymmetric non-Abelian
gauge theories,'' Nucl. Phys. B{\bf 435} (1995) 129--146,\ {\urlfont
hep-th/9411149}.} 

\lref\gh{
P. Grifiths and J. Harris, {\it Principles of Algebraic Geometry}, 
Wiley, 1978.}

\lref\hart{
R. Hartshorne, {\it Algebraic Geometry}, Springer, 1977.}

\lref\ks{
I.~R.~Klebanov and M.~J.~Strassler,
``Supergravity and a Confining Gauge Theory: Duality Cascades and $\chi$SB-Resolution of Naked Singularities,'' J. High Energy Phys. {\bf 08} (2000) 052, 
{\urlfont hep-th/0007191}.}

\lref\kss{
D. Kutasov, A. Schwimmer, and N. Seiberg, ``Chiral Rings, Singularity
Theory, and Electric-Magnetic Duality,'' Nucl. Phys. B {\bf 459}
(1996) 455--496, {\urlfont hep-th/9510222}.}

\lref\kw{
I.~R. Klebanov and E. Witten, 
``Superconformal field theory on threebranes at a
Calabi-Yau singularity,'' Nucl. Phys. B {\bf 536} (1998) 199--218,
{\urlfont hep-th/9807080}.}

\lref\ls{
R~.G. Leigh and M. J. Strassler, ``Exactly marginal operators and
duality in four dimensional $\CN=1$ supersymmetric gauge theory,''
Nucl. Phys. B {\rm 447} (1995) 95--136, {\urlfont hep-th/9503121}.}

\lref\kehagias{
A. Kehagias, ``New type IIB vacua and their F-theory interpretation,''
Phys. Lett. B{\bf 435} (1998) 337--342, {\urlfont hep-th/9805131}.}

\lref\mp{
David R. Morrison and M. Ronen Plesser, 
``Non-Spherical Horizons, I,'' Adv. Theor. Math. Phys. {\bf 3} (1999)
1--81, {\urlfont hep-th/9810201}.}

\lref\svs{
M.~A. Shifman and A.~I. Vainshtein, ``Solution of the anomaly puzzle
in SUSY gauge theories and the Wilson operator expansion,''
Nucl. Phys. B {\rm 277} (1986) 456-486.}

\lref\svh{
M.~A. Shifman and A.~I. Vainshtein, ``On holomorphic dependence and
infrared effects in supersymmetric guage theories,'' Nucl. Phys. B
{\rm 359} (1991) 571--580.}

\lref\ty{
G. Tian and S.-T. Yau, ``K\"ahler-Einstein metrics on complex surfaces
with $C_1 > 0$,'' Comm. Math. Phys. {\bf 112} (1987) 175--203.}

\lref\t{
G. Tian, ``On Calabi's conjecture for complex surfaces of positive
first Chern class,'' Invent. Math. {\bf 101} (1990) 101--172.}

\lref\w{
E. Witten, ``Baryons and branes in anti-de Sitter space,'' J. High
Energy Phys. {\bf 07} (1998) 006, {\urlfont hep-th/9805112}.}

\lref\adscft{See O. Aharony, S.S. Gubser, J. Maldacena, H. Ooguri, and
Y. Oz, ``Large N field theories, string theory and gravity,''
Phys.Rept. {\bf 323} (2000) 183-386, {\urlfont hep-th/9905111}.}

\lref\derived{M.R. Douglas, ``D-branes, categories, and $\CN=1$
supersymmetry,'' {\urlfont hep-th/0011017}.}

\newsec{Introduction.}

The low-energy worldvolume dynamics of a system of $N$ D3-branes is
described by an $\CN=4$ supersymmetric $U(N)$ Yang-Mills theory.  This has a
holographic dual -- a compactification of the type-IIB
string on $AdS_5\times S^5$ \adscft .  At extremely low energies the
worldvolume theory depends on the closed-string backgrounds only
through their values on the branes.  More interesting worldvolume
theories can be obtained by studying sets of intersecting branes or by 
locating the branes near a singularity of the transverse space, and
understanding these theories is of considerable interest. For example, they
shed light on the way in which branes probe spacetime \greene , as well as
providing local models for studying the spectrum of brane states in
nontrivial backgrounds \derived . 

If the singularity in question can be realized by partially resolving
a quotient singularity, then an array of machinery exists to derive
the worldvolume gauge theory governing the low-energy dynamics.  The
theories arising at quotient singularities were first described in \dm
.  To describe $N$ branes near a singularity of the form
$\BC^n/\Gamma$ one begins with $|\Gamma|N$ branes on the covering
space, lifting the action of the discrete group to the Chan-Paton
indices using the regular representation.  The worldvolume theory on
the quotient is given by a projection to $\Gamma$-invariant degrees of
freedom.  One finds a theory with a product gauge group, with factors
associated to the irreducible representations of $\Gamma$, a
supersymetry determined by the $\Gamma$-invariant spinors of $SO(2n)$,
and matter in various representations.  In the cases of interest here,
$n=3$, $\Gamma\subset SU(3)$ preserves four supercharges, the factors
in the unbroken gauge group will be $U(N)$ groups, and the matter
fields will be chiral multiplets in bifundamental representations.
The closed-string background determines the parameters of the theory.
In particular, the metric deformations corresponding to resolutions of
the singularity determine Fayet-Iliopuolos terms in the gauge
theory.\foot{We are being imprecise here. In fact, the worldvolume
gauge group is a product of $SU(N)$ factors for reasons discussed
below, and does not admit FI terms.  The relevant twisted
closed-string modes correspond to expectation values of ``baryonic''
degrees of freedom in the worldvolume theory. The description given
here is nonetheless a useful way of computing the moduli space of
vacua \mp , as we explain.} The moduli space of vacua of the gauge
theory should describe (at least) the positions of the branes in the
transverse space and so contain its $N$-th symmetric power.

In \dgm\ it was noted that for $N=1$ this procedure yields a moduli
space of vacua described naturally as a toric variety.\foot{This is not as
obvious as it sounds; that the moduli space of vacua of an $\CN=1$
supersymmetric Abelian gauge theory is toric is clear.  There are two
subtleties here, however, and these are important for our study.
First, the gauge theory corresponding to the quotient features a cubic
superpotential (descended from the interaction in the $\CN=4$ theory
corresponding to the covering space); the moduli space of vacua is
thus obtained naturally as a subspace of a toric variety given by the
critical points of this.  The insight of \dgm\ is that this space is
itself a toric variety.  A further subtlety, physical rather than
technical, is that the gauge group factors are in fact of the form
$SU(N)$ rather than $U(N)$.  The spaces we will describe are thus not
the full moduli space of vacua but in fact describe a piece of this
associated to motion of the branes in the transverse space.}
This idea was used in \mp\ to systematically discuss the low-energy
dynamics of branes near any toric singularity, and to study an
extension of the holographic conjectures to singular transverse
spaces.  The method used realizes the singularity of interest as a
partial resolution of a $\BZ_n\times\BZ_m$ quotient.  The description
of the quotient theory and the map of metric deformations to FI
parameters is then used to find a theory describing the partially
resolved space.  In essence the FI terms lead to a nonzero expectation
value for some charged fields, leading to spontaneous breaking of the
gauge symmetry.  The extreme low-energy behavior near an
appropriately chosen point in the moduli space of vacua of this theory
then describes the dynamics of the branes near the singularity in
question.  In particular, the low-energy limit is completely
independent of any global properties of the transverse space, as
usual.  By choosing the scale of the blowups (FI terms) appropriately,
it is possible to arrange a situation in which the dynamics below this
scale is determined to a good approximation by the quotient gauge
theory (string corrections are suppressed).  

For a weakly-coupled string theory the dynamics of the theory at the
FI scale is perturbative and the gauge symmetry breaking can be
described by a classical Higgs calculation.  The resulting gauge
theories are in general not conformal.  The expectation is that the
extreme low-energy behavior of these theories will be described by a
nontrivial RG fixed point.  For generic values of the FI terms, of
course, this will simply be the $\CN=4$ theory describing branes at a
smooth point.  At the special values for which we obtain a particular
singular space we should find new conformal theories.  The theory
obtained in this way corresponding to a conical singularity (for which
the cone $C(H)$ over a five-manifold $H$ -- the ``non-spherical
horizon'' -- is a linear approximation) is expected to be
holographically dual to a compactification of the type-IIB string on
$AdS_5\times H$ \mp.  The $\CN=2$ unbroken supersymmetry
of this compactification implies
that $H$ be an Einstein-Sasaki space \kehagias .  In particular, it
must have a $U(1)$ isometry corresponding to the $R$-symmetry of the
worldvolume theory. When the action is regular, the quotient $X$ is a
K\" ahler-Einstein manifold of positive curvature, and $H$ is
recovered as the unit circle bundle in a Hermitian line bundle over
$X$.  The cone $C(H)$ is then the complex cone over $X$.

Toric methods are thus a powerful tool for probing the dynamics of
D3-branes at orbifold singularities and partial resolutions thereof.
When the methods of \mp\ were extended to a more complicated example
in \ceb\ it was found that in general there is a puzzling ambiguity.
The translation of metric deformations to FI terms is far from unique.
Typically, one finds many, many separate cones in the space of
Fayet-Iliopoulos terms which all correspond to the same geometric
phase.  This was systematically studied in \bglp, \fhhtd .  In
general, following the procedure outlined above in the different cones
will lead to {\it distinct\/} gauge theories describing the low-energy
dynamics.  We stated above that the extreme low-energy behavior is
determined completely by the local structure of the singularity, and
the multiplicity of models thus poses a puzzle.  The authors of
\fhhtd\ conjectured that in fact the various models must all flow to
the same infrared conformal fixed point determined by the singularity,
and termed the relation between them ``toric duality.''  There is not,
however, an independent verification of the statement.

In this paper, we study some examples of toric dualities, including
the examples of \fhhtd .  We find that in all these examples, the
various gauge theories corresponding to a given singularity are
related by a version of Seiberg duality \sem , providing a relation between
them intrinsic to the gauge theory.  This is believed to be an exact
statement relating pairs of SQCD theories.  In our models, the SQCD in
question is deformed by coupling to other degrees of freedom in the
theory, and when we say two models are related by Seiberg duality we
mean they are given by dual deformations of the two dual SQCD theories.

In general, the gauge theories we find are not asymptotically free.
This means we cannot usefully think of the RG trajectory we follow to
the conformal fixed point in terms of the weakly coupled theory.
Technically, the gauge couplings for these factors are an example of
``dangerously irrelevant'' deformations.  In fact, even when our
models are asymptotically free, the superpotentials we find contain
such operators.  Theories containing such operators were studied in a
related context in \kss , and our models exhibit many of the
subtleties found there.  For instance, we find that a perturbative calculation
of anomaly coefficients in each of the four models does not satisfy the
anomaly-matching criterion of 't Hooft.  We also find that the classical 
rings of chiral operators in the four models are truncated by additional 
quantum relations in the infrared, a statement which we motivate both 
from Seiberg duality and the \AdS/CFT correspondence.

We will focus here on understanding a particular example of toric
duality and will indicate the extension to the other known examples.
The example of interest arises from the partial resolution of 
$\BC^3/(\BZ_3 \times \BZ_3)$ which is locally a complex
cone over a del Pezzo surface $X= dP_3$ which can be realized as 
a blowup of $\BP^2$ at three generic points.  $X$ is known to possess 
a smooth K\"ahler-Einstein metric \ty,\t\ and the corresponding 
horizon $H$ is Einstein-Sasaki with regular $U(1)$-action \mp.  In 
this case, there are in fact 1,602 cones in the space of 
Fayet-Iliopoulos parameters corresponding to the partial 
resolution which leads to this singularity.  Values of the 
FI parameters in each of these cones lead to one of four distinct 
$\CN=1$ $SU(N)^6$ gauge-theories whose infrared
dynamics describe the low-energy excitations of $N$ coincident
D3-branes transverse to this singularity\foot{As a sidenote, we remark
that this fact casts doubt on the explanation offered in \fhhps\ for
the toric duality, since the authors' explanation would predict a
unique gauge theory for the case $X=dP_3$.  In fact, the various
global choices in how one resolves the orbifold (to which the authors
attribute the existence of the distinct gauge theories) are related
by the large quantum symmetry group of the $\BZ_3 \times \BZ_3$
orbifold gauge theory and so cannot be responsible for the
ambiguity.}.  The presence of the $\CN=1$ supersymmetry makes the
study of the infrared behaviour of these gauge theories a tractable
problem.

In Section 2, we first sketch the $\BC^3/(\BZ_3 \times \BZ_3)$
orbifold gauge theory and derive from it the four gauge theories
describing the complex cone over $X$.  We study the four theories in
some detail, finding evidence to support the conjecture that they flow
to a common fixed point. We then demonstrate that the four gauge
theories are related to each other by a sequence of $\CN=1$ dualities.

In Section 3, we review the classical algebraic geometry and topology
of $X$ and the cone $C$, and the differential geometry of the non-spherical
horizon $H$ associated to the dual
\AdS\ description.  This analysis is very useful for understanding
some general properties of the four gauge theories.  It is
necessary to understand and explain the spectrum of baryonic operators
in the gauge theories via wrapped D3-brane states in the dual \AdS\
theory.  (Unfortunately, a rather detailed analysis of the geometry is
required for this purpose, but we believe that the payoff for this
effort in Section 4 makes it worthwhile.)

In Section 4, we apply the $\CN=1$ dualities to understand quantum
corrections to the ring of chiral operators in the gauge theories.  We
use this determination of the algebra of chiral operators to describe
the spectrum of D3-branes wrapping supersymmetric three-cycles on $H$ in the 
dual \AdS\ theory.

In Section 5, we discuss the other known examples of toric duality
\fhhtd\ and show how they, too, can be understood using Seiberg
duality.  We conclude in Section 6 with some ideas for further study.

In the Appendix, we list some relevant facts about chiral operators in
each of the four gauge theories.

\newsec{The four \dP\ models.}

As a review, we first recall how the four gauge theories describing
$N$ D3-branes transverse to the cone over \dP\ arise from the
$\BC^3/(\BZ_3 \times \BZ_3)$ orbifold theory.  The orbifold theory
itself is determined by lifting $N$ coincident D3-branes at the origin
of the orbifold to $9N$ coincident D3-branes on the cover.  The
worldvolume $U(9N)$ super Yang-Mills theory on the covering branes is
then projected onto the sector invariant under the $\BZ_3 \times
\BZ_3$ action, lifted to the Chan-Paton bundle using the regular
representation.  Projecting to the invariant fields reduces the
$U(9N)$ gauge group to $U(N)^9$.  The $U(1)$ factors (except the
diagonal ``center of mass'' factor, which decouples) are broken by the
coupling to twisted sector closed-string states \dm .  The gauge
group on the worldvolume is thus $SU(N)^9$, while the $U(1)^8$ remains a
global symmetry of the theory.  The orbifold preserves an $\CN=1$
supersymmetry.  The superpotential of the orbifold theory is the
projection of the $\CN=4$ superpotential $W = \tr ([X,Y]Z)$ and hence
is also cubic.  All chiral matter transforms in bifundamental
representations of the gauge group, and the representation content of
the theory is neatly summarized by a quiver diagram (see \bglp\ for
this quiver).

We will also use quiver diagrams to summarize the representation
contents of the models of interest, so we mention how these diagrams are
read.  In quiver diagrams, the nodes represent the semi-simple $SU(N)$
factors of the $SU(N)^p$ gauge-group, and an arrow from node $i$ to
node $j$ indicates a chiral superfield which transforms as $\bkb$
under $SU(N)_i$, as $\bk$ under $SU(N)_j$, and as a singlet under the
other factors of the gauge-group.  Such a field we denote by $X_{ji}$.

In this notation, the matter content of the $\BC^3/(\BZ_3 \times
\BZ_3)$ orbifold theory consists of twenty-seven chiral superfields
\eqn\Xs{\eqalign{
&X_{15}, X_{26}, X_{34}, X_{48}, X_{59}, X_{67}, X_{72}, X_{83}, X_{91},\cr
&Y_{13}, Y_{21}, Y_{32}, Y_{46}, Y_{54}, Y_{65}, Y_{79}, Y_{87}, Y_{98},\cr
&Z_{17}, Z_{28}, Z_{39}, Z_{41}, Z_{52}, Z_{63}, Z_{74}, Z_{85}, Z_{96},\cr
}}
and the orbifold superpotential is
\eqn\W{\eqalign{
W &= \tr\Big[
Z_{17}(X_{72}Y_{21} - Y_{79}X_{91}) +
Z_{28}(X_{83}Y_{32} - Y_{87}X_{72}) + \cr
&\qquad Z_{39}(X_{91}Y_{13} - Y_{98}X_{83}) +
Z_{41}(X_{15}Y_{54} - Y_{13}X_{34}) + \cr
&\qquad Z_{52}(X_{26}Y_{65} - Y_{21}X_{15}) +
Z_{63}(X_{34}Y_{46} - Y_{32}X_{26}) + \cr
&\qquad Z_{74}(X_{48}Y_{87} - Y_{46}X_{67}) +
Z_{85}(X_{59}Y_{98} - Y_{54}X_{48}) + \cr
&\qquad Z_{96}(X_{67}Y_{79} - Y_{65}X_{59})\Big]\ .\cr}}

The moduli space of classical vacua of this theory will be given by
the critical points of $W$ in the symplectic quotient of field space
by the $SU(N)^9$ action.  This space describes deformations of the
theory by expectation values for all worldvolume scalars, including
both open string states (describing motion of the branes in the
transverse space) and twisted closed string states (describing
deformations of the orbifold background).  These can be separated by
the following useful device \mp .  The baryonic $U(1)^8$ global
symmetry acts on the space of classical vacua, preserving the
symplectic structure.  The symplectic quotient of the moduli space by
this symmetry, determined by a moment map, yields a space
corresponding to the theory with gauge group $U(N)^9/U(1)$.  This
``mesonic'' moduli space describes motion of the branes in a
deformation of the orbifold determined by the moment map, which here
appears as a Fayet-Iliopoulos term.  The full moduli space thus
fibers over the space $\BR^8$ of moment map values, with the fiber
over each point a $U(1)^8$ bundle over the corresponding mesonic moduli
space.  The degrees of freedom parameterizing this extra structure are
the ``baryonic'' states charged under $U(1)^8$, and describe the
deformations of the orbifold background by twisted closed string
states.  We will thus find it useful to speak about the theory in
terms of Fayet-Iliopoulos terms, keeping in mind that we are in fact
describing baryonic moduli.

In the orbifold gauge theory, a non-zero Fayet-Iliopoulos term implies
that, in order to satisfy the corresponding $D$-term equation and
preserve supersymmetry, some chiral superfields charged under the
corresponding $SU(N)$ factor must have a non-zero vacuum
expectation-value.  These vevs generically break some or all of the
$SU(N)^9$ gauge group and introduce a scale into the (previously
conformal) orbifold theory.  Because of the cubic superpotential \W,
these vevs also generically induce mass terms for some of the
chiral matter.  So the resulting Higgsed theory is not conformal, but
one assumes that upon renormalization-group flow to the infrared one
reaches a conformal point describing the new geometric phase.  If the
string theory is weakly coupled, the symmetry breaking and integrating
out of massive fields can reliably be performed perturbatively.

The work of \bglp\ implies that the four gauge theories describing the
cone $C$ over \dP\ can be realized by Fayet-Iliopoulos deformations which
have the effect of turning on large, generic vevs for the following
four sets of chiral fields in the orbifold theory:
\eqn\vevs{\matrix{
&(I):&X_{83},&Y_{79},&Z_{41}, \cr 
&(II):&X_{67},&Y_{87},&Z_{53}, \cr
&(III):&X_{15},&Y_{32},&Z_{17}, \cr
&(IV):&X_{26},&Y_{21},&Z_{28}. \cr}} 
The fact that each set of vevs involves an $X$, $Y$, and $Z$ chiral
field follows from the fact that geometrically a $\BZ_3$ subgroup of
the symmetry group of the orbifold is realized as a symmetry of $C(H)$.
However, it is interesting to note that, in terms of
the orbifold gauge theory, the sets of vevs in \vevs\ break differing
amounts of the global symmetry.  As we shall see, the
effective gauge theories we obtain will be invariant under different
global symmetry groups.

Given the expectation values \vevs, we can classically integrate out
the massive fields to obtain the effective descriptions at energies
below the Higgs scale.  The expectation is that all of these will flow
in the extreme infrared to the same conformal field theory.
Each model has $\CN=1$ supersymmetry, a gauge group
$SU(N)^6$, a number of chiral superfields transforming as
bi-fundamentals under the gauge group, canonical K\"ahler terms, and a
superpotential.  
\midinsert
$$\matrix{
&{\epsfxsize=2in\epsfbox{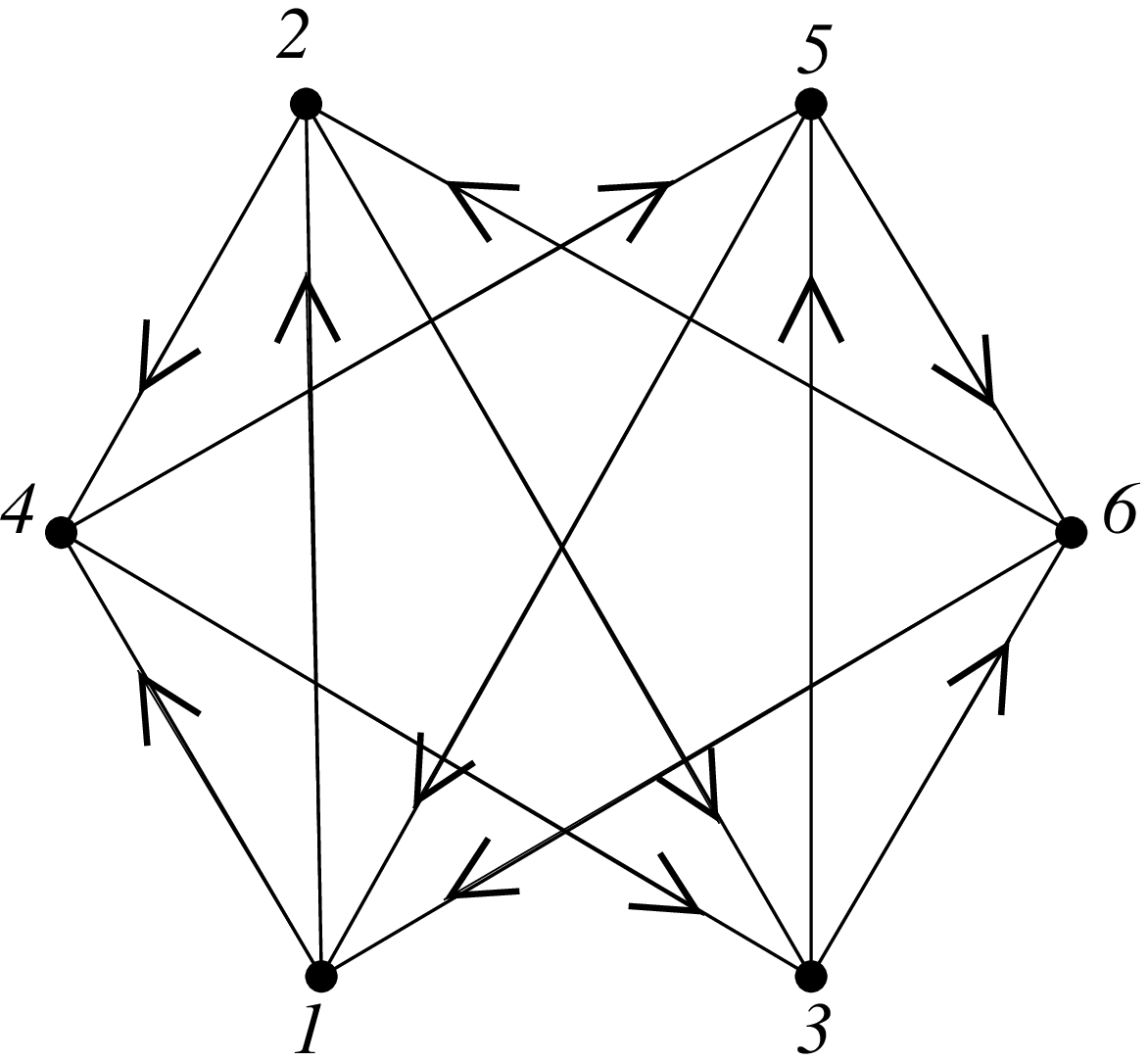}}\qquad
&{\epsfxsize=2in\epsfbox{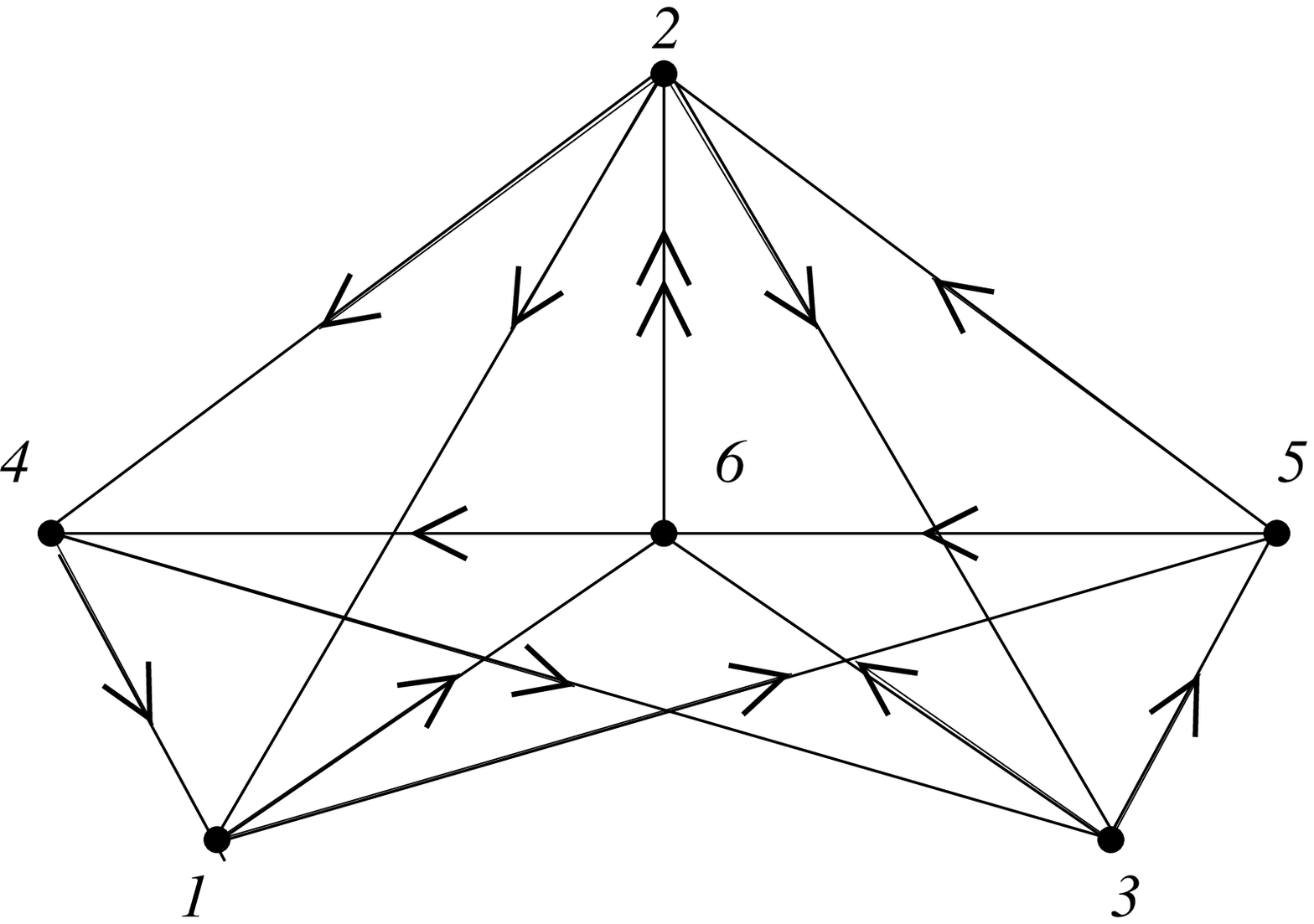}}\cr
&\hbox{Model I}\qquad&\hbox{Model II}\cr\cr\cr
&{\epsfxsize=2in\epsfbox{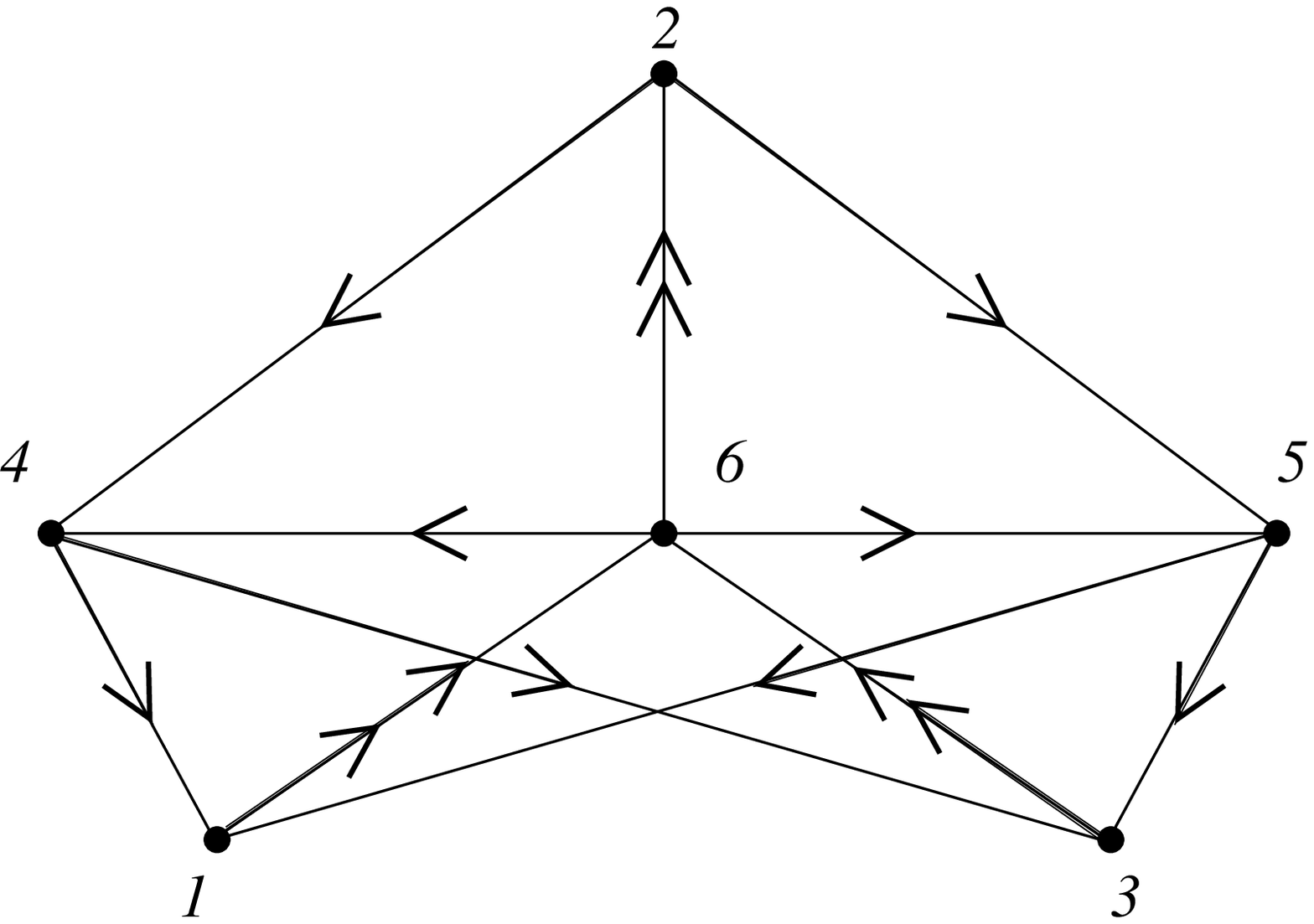}}\qquad
&{\epsfxsize=2in\epsfbox{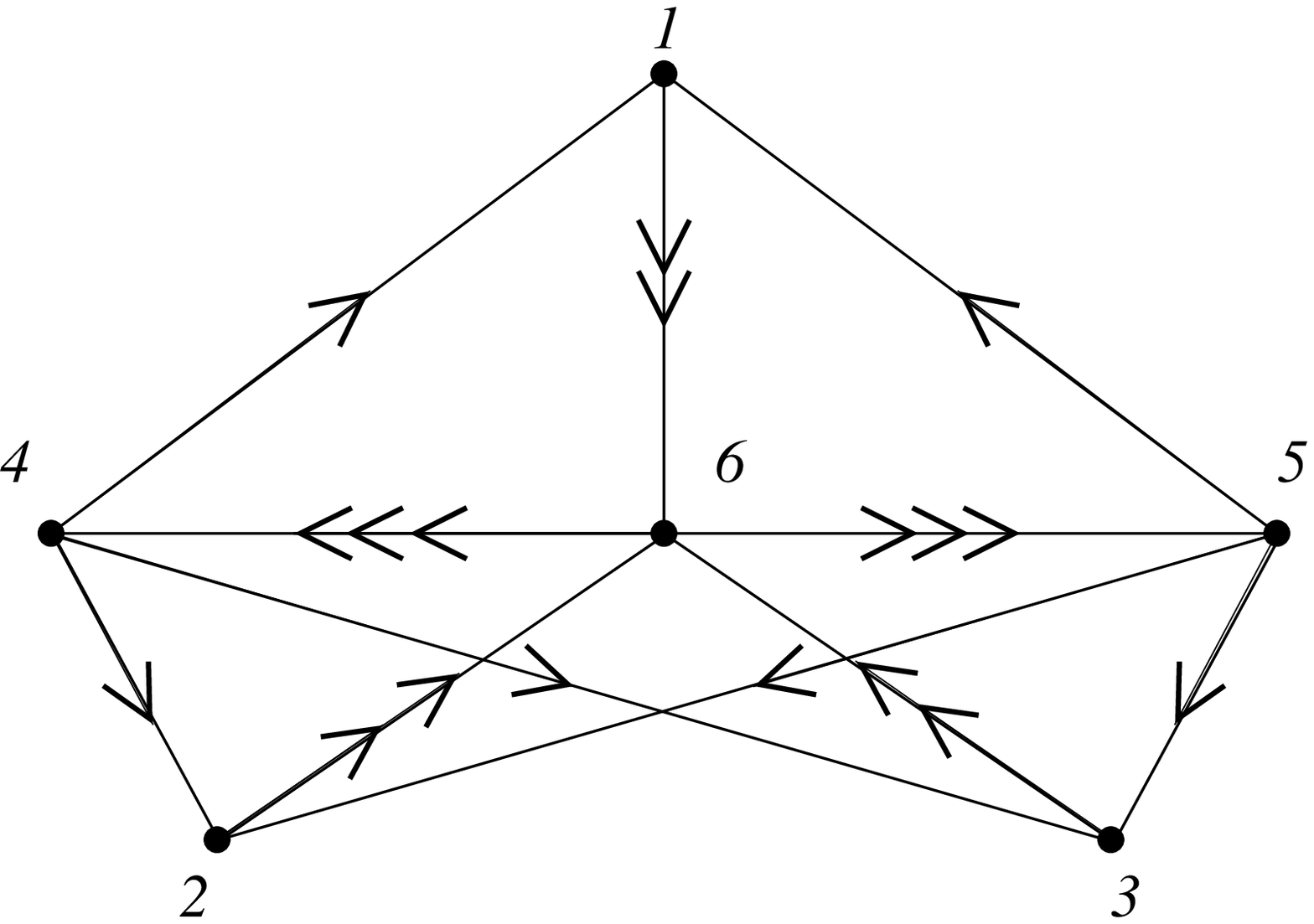}}\cr
&\hbox{Model III}\qquad&\hbox{Model IV}}$$
\smallskip
\centerline{Fig. 1. Quivers of the four models.}
\bigskip
\endinsert
The representation contents of the four theories (to which
we refer as Model I, Model II, Model III, and Model IV) are summarized
by the quiver diagrams in Figure 1.  For each of the four models, we
also list below the classical superpotentials 
(which arise from the orbifold superpotential
\W\ after integrating-out massive fields):
\eqn\wi{\eqalign{
W_{I} &= h_1 \tr \Big[X_{32} X_{26} X_{63}\Big] + h_2 \tr\Big[X_{41}
X_{15} X_{54}\Big] + h_3 \tr\Big[X_{34} X_{42} X_{21} X_{16} X_{65}
X_{53}\Big] \cr 
&- h_4 \tr\Big[X_{41} X_{16} X_{63} X_{34}\Big] - h_5
\tr\Big[X_{21} X_{15} X_{53} X_{32}\Big] - h_6 \tr\Big[X_{54} X_{42}
X_{26} X_{65}\Big]\,,}}
\smallskip
\eqn\wii{\eqalign{
W_{II} &= h_1 \tr\Big[Y_{46} Y_{61} X_{14}\Big] + h_2 \tr\Big[X_{25}
X_{51} X_{12}\Big] - h_3 \tr\Big[Y_{46} X_{63} X_{34}\Big] - h_4
\tr\Big[X_{25} X_{53} X_{32}\Big] \cr 
&+ h_5 \tr\Big[X_{32} X_{26}
X_{63}\Big] - h_6 \tr\Big[X_{12} Y_{26} Y_{61}\Big] \cr &+ h_7
\tr\Big[X_{34} X_{42} Y_{26} X_{65} X_{53}\Big] - h_8 \tr\Big[X_{14}
X_{42} X_{26} X_{65} X_{51}\Big]\,,}}
\smallskip
\eqn\wiii{\eqalign{
W_{III} &= h_1 \tr\Big[ X_{56} X_{61} X_{15}\Big] - h_2 \tr\Big[X_{56}
Y_{63} X_{35} \Big] + h_3 \tr\Big[ Y_{46} Y_{61} X_{14} \Big] - h_4
\tr\Big[ Y_{46} X_{63} X_{34} \Big] \cr
&+ h_5 \tr\Big[ X_{42} Y_{26} Y_{63} X_{34} \Big] - h_6 \tr\Big[
X_{52} Y_{26} Y_{61} X_{15} \Big] \cr &+ h_7 \tr\Big[ X_{52} X_{26}
X_{63} X_{35} \Big] - h_8 \tr\Big[ X_{42} X_{26} X_{61} X_{14}
\Big]\,,}}
\smallskip
\eqn\wiv{\eqalign{
W_{IV} &= h_1 \tr \Big[ Y_{56} X_{63} X_{35} \Big] - h_2 \tr
\Big[Y_{56} Y_{62} X_{25} \Big] + h_3 \tr \Big[ Y_{46} Y_{61} X_{14}
\Big] - h_4 \tr \Big[ Y_{46} X_{63} X_{34} \Big] \cr
&+ h_5 \tr \Big[ X_{56} X_{61} X_{15} \Big] - h_6 \tr \Big[ X_{56}
Y_{63} X_{35} \Big] + h_7 \tr \Big[ X_{46} Y_{62} X_{24} \Big] - h_8
\tr \Big[ X_{46} X_{61} X_{14} \Big] \cr &+ h_9 \tr \Big[ Z_{56}
X_{62} X_{25} \Big] - h_{10} \tr \Big[ Z_{56} Y_{61} X_{15} \Big] +
h_{11} \tr \Big[ Z_{46} Y_{63} X_{34} \Big] - h_{12} \tr \Big[ Z_{46}
X_{62} X_{24} \Big]\,.}}
\smallskip

The parameters $h_{i}$ are (in general dimensionful) couplings,
determined classically in terms of the (generically chosen)
expectation values \vevs .  The $\CN=1$ supersymmetry ensures that the
classical superpotentials will not be renormalized in perturbation
theory except via wavefunction renormalization.  Varying $h_i$ will
correspond to marginal or irrelevant deformations of the theory at
weak coupling; we assume this holds in general and thus that a choice
of these comprises at most a choice of a point on the critical
manifold.  We will choose the (nonzero) coefficients to maximize the
unbroken classical global symmetry of the model.  

We now proceed to study these models and present the evidence relating
them to one another as well as 
degree to which we can demonstrate that they reproduce the properties
expected from the geometry of $X$ and $H$ (to be discussed in the next
section). 

\subsec{Chiral operators.}

We first compare the algebras of chiral operators in the four models.
In any quiver theory containing only bi-fundamental chiral
fields, the gauge-invariant operators fall into two classes.  First,
for each chiral operator $\CO$ which transforms under the various
factors of the gauge-group as an adjoint (or a singlet), we have the
gauge-invariant operator $\tr[\CO]$.  We refer to these operators as
mesonic operators.  Second, since only $SU(N)$ factors occur in the
gauge-group (the $U(1)$ factors of a naive $U(N)$ are 
broken as discussed above), we may use the invariant anti-symmetric
tensor to make baryonic gauge-invariant chiral operators.  If $\CO$ is
a chiral operator transforming as a fundamental or anti-fundamental
(or singlet) under each $SU(N)$ factor, then we form a gauge-invariant
baryonic operator by anti-symmetrizing on each gauge-index, i.e. in
the case of a bi-fundamental operator, $\epsilon_{i_1 \ldots i_N}
\epsilon^{j_1 \ldots j_N} (\CO)^{i_1}_{j_1} \ldots (\CO)^{i_N}_{j_N}$.
We abbreviate such an operator by $\CO^N$.  In general, we may have
multiple chiral operators $\CO_1$, $\CO_2$, etc. transforming in the
same representation.  For simplicity, assuming $\CO_1$ and $\CO_2$ are
both bi-fundamentals, then we may also make baryonic operators of the
form $\epsilon_{i_1 \ldots i_N} \epsilon^{j_1 \ldots j_N}
(\CO_1)^{i_1}_{j_1} \ldots (\CO_1)^{i_s}_{j_s}
(\CO_2)^{i_{s+1}}_{j_{2+1}} \ldots (\CO_2)^{i_N}_{j_N}$.
We abbreviate such operators by $\CO_1^s \CO_2^{N-s}$, and
similarly for baryonic operators involving more than two flavors.

As discussed above, the baryonic operators are associated to
deformations of the closed string background by twisted fields.  The
mesonic operators, on the other hand, will describe motion of the
branes.  In terms of the holographically dual \AdS\ dual, the baryonic
operators, which carry $\CR$-charge ( and hence conformal weight)
proportional to $N$, are associated to stringy effects, and the
baryons themselves correspond to wrapped branes. Hence we first
consider the mesonic operators.  The chiral ring is (classically)
isomorphic to the ring of gauge-invariant holomorphic functions in the
fields modulo the Jacobian ideal of $W$.  In each of the theories the
mesonic ring is generated by seven operators $\CM_i$.  In the
appendix, we list these as functions of the fields for each model.

A first check of the claimed duality is that the subalgebras of the
chiral rings in each of the four models which are generated by the
mesonic invariants agree (and do realize the geometry of the cone
$C$).  It is easy, if somewhat tedious, to check that in each of the
four models all relations between the mesonic generators $\CM_i$ are
generated by the following nine quadratic polynomials:
\eqn\rel{\eqalign{
&\CM_1 \CM_7 - \CM_2 \CM_6 \, , \quad \CM_2 \CM_7 - \CM_1 \CM_3 \, ,
\quad \CM_3 \CM_7 - \CM_2 \CM_4 \, , \cr 
&\CM_4 \CM_7 - \CM_3 \CM_5 \, , \quad \CM_5 \CM_7 - \CM_4 \CM_6 \, , 
\quad \CM_6 \CM_7 - \CM_1 \CM_5\, , \cr 
&\CM_7^2 - \CM_1 \CM_4 \; \, , \quad \CM_7^2 - \CM_2 \CM_5 \;
\, , \quad \CM_7^2 - \CM_3 \CM_6. \; \, \cr}} 

As we will show in Section 3, the ideal generated by these nine
polynomials also presents the cone $C$ as a subvariety of $\BC^7$
(identifying the coordinates of $\BC^7$ with the mesonic generators
$\CM_i$ in the above).  So the invariant traces of products of the
mesonic generators parametrize the $S_N$ invariants of $N$ points
satisfing \rel, and we identify the mesonic branch of the moduli space
as the symmetric product of $N$ copies of the cone $C$ (as we expect
for $N$ identical branes placed transverse to $C$).  Our investigation of
the mesonic subalgebra of the chiral ring is thus consistent 
with the conjecture that the models are ``dual'' and flow to the 
same infrared fixed point.

\subsec{Global symmetries.}

We now examine the global symmetries in each of the four models.  The
continuous global symmetries are Abelian.  Each of the models has a
nonanomalous $U(1)^6$ global symmetry, and the charges of the chiral
superfields under this are listed in Table 1.  In this table we have
chosen a basis for this $U(1)^6$ in each model, and we have labelled
the individual factors as $U(1)_A$, $U(1)_B$, $U(1)_C$, $U(1)_D$,
$U(1)_E$, and $U(1)_\CR$.  The last of these, of course, acts as an
$R$-symmetry.  The identification of these bases between models is
nontrivial.  The mesonic operators in all four theories are charged
only under the subgroup $U(1)_A \times U(1)_B \times U(1)_\CR$, with
charges
\eqn\uabr{\pmatrix{
&\CM_1&\CM_2&\CM_3&\CM_4&\CM_5&\CM_6&\CM_7\cr 
&0&-1&-1&0&1&1&0\cr
&1&1&0&-1&-1&0&0\cr &2&2&2&2&2&2&2\cr}\,. } 
We will see that the action of this $U(1)^3$ on the
moduli space in the gauge theory is in agreement with an expected
$U(1)^3$ isometry of the horizon.  Under the additional $U(1)_C \times
U(1)_D \times U(1)_E$ symmetries, only the baryonic operators in each
model are charged.  Such a baryonic symmetry corresponds in the dual
\AdS\ theory to $U(1)$ gauge symmetries arising from the reduction of
the R-R four-form about three-cycles on $H$.  We will discuss these cycles
in Section 3 and will find agreement with this prediction.  In Section
4, we will return to this symmetry as we identify baryons with wrapped
D3-branes.
\midinsert
$$\vbox{\offinterlineskip\halign{
\strut # height 125pt depth 125pt&
\quad$#$\quad\hfill\vrule&\quad$#$\quad\hfill\vrule\cr
\noalign{\hrule}
\vrule &
\matrix{
\hbox{Model I}&&&&&& \cr
&A&B&C&D&E&\CR \cr
X_{21}& 1& 1& 1& 0& 1&1/3 \cr
X_{65}& 0& 0&-1&-1& 1&1/3 \cr
X_{34}& 0& 0& 0& 1& 1&1/3 \cr
X_{53}& 0& 1& 1& 0&-1&1/3 \cr
X_{42}& 0&-1&-1&-1&-1&1/3 \cr
X_{16}&-1&-1& 0& 1&-1&1/3 \cr
X_{32}&-1&-1&-1& 0& 0&2/3 \cr
X_{15}& 0&-1&-1& 0& 0&2/3 \cr
X_{26}& 0& 1& 1& 1& 0&2/3 \cr
X_{54}& 0& 0& 1& 1& 0&2/3 \cr
X_{63}& 1& 0& 0&-1& 0&2/3 \cr
X_{41}& 0& 1& 0&-1& 0&2/3 \cr
\cr \cr \cr \cr \cr \cr} &
\matrix{
\hbox{Model II}&&&&&& \cr
&A&B&C&D&E&\CR \cr
X_{14}&  1&  1&  1&  0&  1&1/3 \cr
X_{65}&  0&  0& -1& -1&  1&1/3 \cr
X_{34}&  0&  0&  0&  1&  1&1/3 \cr
X_{53}&  0&  1&  1&  0& -1&1/3 \cr
X_{42}&  0& -1& -1& -1& -1&1/3 \cr
X_{51}& -1& -1&  0&  1& -1&1/3 \cr
X_{32}& -1& -1& -1&  0&  0&2/3 \cr
Y_{61}&  0& -1& -1&  0&  0&2/3 \cr
X_{26}&  0&  1&  1&  1&  0&2/3 \cr
Y_{26}&  0&  0&  1&  1&  0&2/3 \cr
X_{63}&  1&  0&  0& -1&  0&2/3 \cr
X_{12}&  0&  1&  0& -1&  0&2/3 \cr
Y_{46}& -1&  0&  0&  0& -1&  1 \cr
X_{25}&  1&  0&  0&  0&  1&  1 \cr
\cr \cr \cr \cr} \cr
\noalign{\hrule}
\vrule &
\matrix{
\hbox{Model III}&&&&&& \cr
&A&B&C&D&E&\CR \cr
X_{14}& 1& 1& 1& 0& 1&1/3 \cr
X_{52}& 0& 0&-1&-1& 1&1/3 \cr
X_{34}& 0& 0& 0& 1& 1&1/3 \cr
X_{15}& 0& 1& 1& 0&-1&1/3 \cr
X_{42}& 0&-1&-1&-1&-1&1/3 \cr
X_{35}&-1&-1& 0& 1&-1&1/3 \cr
X_{61}&-1&-1&-1& 0& 0&2/3 \cr
Y_{61}& 0&-1&-1& 0& 0&2/3 \cr
X_{26}& 0& 1& 1& 1& 0&2/3 \cr
Y_{26}& 0& 0& 1& 1& 0&2/3 \cr
X_{63}& 1& 0& 0&-1& 0&2/3 \cr
Y_{63}& 0& 1& 0&-1& 0&2/3 \cr
Y_{46}&-1& 0& 0& 0&-1& 1 \cr
X_{56}& 1& 0& 0& 0& 1& 1 \cr
\cr \cr \cr \cr} &
\matrix{
\hbox{Model IV}&&&&&& \cr
&A&B&C&D&E&\CR \cr
X_{14}& 1& 1& 1& 0& 1&1/3 \cr
X_{24}& 0& 0&-1&-1& 1&1/3 \cr
X_{34}& 0& 0& 0& 1& 1&1/3 \cr
X_{15}& 0& 1& 1& 0&-1&1/3 \cr
X_{25}& 0&-1&-1&-1&-1&1/3 \cr
X_{35}&-1&-1& 0& 1&-1&1/3 \cr
X_{61}&-1&-1&-1& 0& 0&2/3 \cr
Y_{61}& 0&-1&-1& 0& 0&2/3 \cr
X_{62}& 0& 1& 1& 1& 0&2/3 \cr
Y_{62}& 0& 0& 1& 1& 0&2/3 \cr
X_{63}& 1& 0& 0&-1& 0&2/3 \cr
Y_{63}& 0& 1& 0&-1& 0&2/3 \cr
X_{46}& 0& 0& 0& 0&-1& 1 \cr
Y_{46}&-1& 0& 0& 0&-1& 1 \cr
Z_{46}& 0&-1& 0& 0&-1& 1 \cr
X_{56}& 1& 0& 0& 0& 1& 1 \cr
Y_{56}& 0& 1& 0& 0& 1& 1 \cr
Z_{56}& 0& 0& 0& 0& 1& 1} \cr
\noalign{\hrule}
}}$$
\centerline{Table 1.  $U(1)^6$ Charges in Models I, II, III, and IV.} 
\bigskip 
\endinsert

Although we will examine the baryonic chiral operators in more detail
in Section 4, we make a few remarks now.  The chiral algebra of
baryonic operators carries a $\BZ$-grading by $\CR$-charge, or
equivalently conformal-weight, and we will restrict our discussion to
the three lowest levels of baryonic operators corresponding to
$\CR$-charge ${1 \over 3}N$, ${2 \over 3}N$, and $N$.  In the Appendix
we list generators $\CB_i$, $\CB_i(s)$, and $\CB_i(s,t)$ of the
classical rings of baryonic operators for these three levels.  Note
that in each of the four models we find the same number of generators
at each level, with corresponding generators having the same charges
under the global $U(1)^6$.  

We now discuss the maximal discrete global symmetries of the four
models (corresponding to the most symmetric choice of the couplings
$h_i$ and gauge-couplings in each model).  As we will explain in
Section 3, the order-12 dihedral group\foot{This group acts as the
symmetries of a regular hexagon and is isomorphic to $S_3 \times
\BZ_2$.} $D_{12}$ acts as a symmetry of $X$, and it also acts as a
symmetry of the non-spherical horizon $H$ and the cone $C$.  So
we expect to see a $D_{12}$ discrete symmetry in the four models.  
In fact, we only see this symmetry in two of the models.

In each model, the discrete symmetries act by permutations of the
$SU(N)$ factors in the gauge group, along with appropriate
permutations of the chiral superfields and possibly
charge-conjugation. So we specify these symmetries by supplying a
cycle-decomposition of these permutations between $SU(N)$ factors
(and also the corresponding action on chiral fields where this is not
uniquely determined by the charges), and we denote
charge-conjugation by $\CC$.  For instance, in Model I we see that the
expected $D_{12}$ symmetry is generated by the following order-6 and
order-2 elements:
\eqn\disi{
(612435), \quad{\rm and}\ \ (45)(26)\CC \, .
}  
On the mesonic generators $\CM_i$ these act as the permutations
\eqn\disim{
(123456), \quad{\rm and}\ \ (26)(35)\, .
}
As we shall see, this agrees precisely with the $D_{12}$ action on $X$
and $H$. 

In the other three models, these discrete symmetries also include an
action by phases on the chiral fields.  These additional phases arise
under those permutations which are $\CR$-symmetries, for which the
superpotential transforms as $W \mapsto -W$.  In order to preserve the
sign of $W$, we must include with these permutations a factor $\exp(i
{\pi \over 2} r)$, where $\exp(i {\pi \over 2} r)$ is the naive
$\CR$-symmetry which acts on the gauginos as $\lambda \mapsto i
\lambda$, on all chiral fermions as $\psi \mapsto -i \psi$, and fixes
the scalar components of the chiral fields.  Generally, though, the
combined permutation and rotation by the naive $\CR$-symmetry will be
anomalous (since the naive $\CR$-symmetry itself is anomalous in these
models).  A similar situation was encountered in \kw, but in that case
the outer-automorphism of the gauge-group was enough to ensure that
the discrete symmetry was non-anomalous.  In these models, in order to
find a non-anomalous discrete symmetry we must explicitly cancel the
anomaly in the naive discrete symmetry with another anomaly arising
from an additional $U(1)_a$ action.  This $U(1)_a$ action manifests
itself as the additional phases on chiral fields to which we alluded
above.

For instance, in Model II we have the following discrete symmetries:
\eqn\disii{
(45)(26)\CC , \quad (13)(X_{26} Y_{26}) 
\exp(i {\pi \over 2} r) \exp(i \pi a)  \, .
} 
First, note that we do not see the full $D_{12}$ as expected, but
only a $\BZ_2 \times \BZ_2$ subgroup.  Second, the symmetry involving
the permutation $(13)$ is an $\CR$-symmetry, for which we have
included the factor $\exp(i {\pi \over 2} r)$.  The naive
$\CR$-symmetry is anomalous, which means that in the presence of $k_j$
instantons in the $j$-th $SU(N)$ factor of the gauge group, the
path-integral measure transforms with a factor of $i$ for each of the
$2 N k_j$ gaugino zero-modes associated to $SU(N)_j$ and a factor of
$-i$ for each of the $N k_j$ zero-modes of each chiral field charged
under $SU(N)_j$.  This anomaly implies that the measure picks up an
overall phase $(-1)^{N (k_1 + k_3 + k_4 + k_5)}$ under $\exp(i {\pi
\over 2} r)$.  Since the outer-automorphism $(13)$ of $SU(N)^6$
exchanges the first and third $SU(N)$ factors, the classical instanton
background which contributes to the anomaly satisfies $k_1 =
k_3$---but we still have a possible anomalous phase arising from
instantons in the fourth and fifth $SU(N)$ factors.  We cancel this
anomaly with the (anomalous) $\BZ_2$ symmetry $\exp (i \pi a)$ under
which the fields $X_{14}$, $X_{34}$, $X_{51}$, $Y_{26}$, $X_{12}$, and
$Y_{46}$ transform by a phase $-1$.  Under this $\BZ_2$, the
zero-modes of each chiral field contribute a phase of $(-1)^{N
(k_1+k_3+k_4+k_5)}$ to cancel the anomaly.  On the mesonic operators
$\CM_i$, this generator then acts as
\eqn\sa{\eqalign{
&\CM_1 \rightarrow - \CM_4\, , \quad \CM_4 \rightarrow - \CM_1 \, ,\cr
&\CM_2 \rightarrow \CM_3\, , \quad \CM_3 \rightarrow - \CM_2 \, ,\cr
&\CM_5 \rightarrow \CM_6\, , \quad \CM_6 \rightarrow - \CM_5 \, , \cr
&\CM_7 \rightarrow \CM_7 \, ,}} 
corresponding to the action of the order-2 element of $S_3$ in \disim\
and a $\pi$-rotation in the second $U(1)$ subgroup of \uabr\ (hence we could
remove the phases above by including with the $\BZ_2$ an action of
this $U(1)$).

In Model III, we also see only a $\BZ_2 \times \BZ_2$ subgroup of
$D_{12}$.  The generators for this symmetry are
\eqn\disiii{
(45)(X_{61} Y_{61})(X_{63} Y_{63})(X_{26} Y_{26})\, , 
\quad (13)(X_{26} Y_{26}) \exp(i {\pi \over 2} r) \exp( i \pi a) \, .
}
Here $\exp (i \pi a)$ acts with a phase of $(-1)$ on the fields
$X_{34}$, $X_{61}$, $X_{63}$, $Y_{63}$, $X_{26}$, and $X_{56}$.  The
action on the mesonic operators $\CM_i$ is as in \sa.

Finally, in Model IV we again see the full $D_{12}$ discrete symmetry.
The order-6 and order-2 elements generating this symmetry are
respectively
\eqn\disiv{\eqalign{
&(123)(45)(X_{56} X_{46} Y_{56} Y_{46} Z_{56} Z_{46})(X_{61} Y_{62}
X_{63} Y_{61} X_{62} Y_{63})\, , \cr
&(12)(X_{56} Y_{56})(Y_{46} Z_{46})(X_{61} Y_{62})(Y_{61} X_{62})
(X_{63} Y_{63}) \exp(i \pi r) \exp(i \pi a)\, .
}}
Here $\exp(i \pi a)$ acts with a phase $(-1)$ on the fields charged
under the fourth $SU(N)$, i.e. $X_{14}$, $X_{24}$, $X_{34}$, $X_{46}$,
$Y_{46}$, and $Z_{46}$.  The $D_{12}$ action on the $\CM_i$ in Model
IV is as in \disim .

Two aspects of the global symmetry of the four models seem to
challenge the assertion that they all flow to the same conformal fixed
point.  The first, that only two of the models exhibit the full
expected discrete symmetry, is not very serious.  The 
$\BZ_3$ generator absent from Model II and III could be an
accidental symmetry which is restored in the infrared.

The second puzzle involves the 't Hooft anomaly matching conditions.
One of the standard checks for duality relations among theories
flowing to identical conformal fixed points involves the equality of
the anomalies in global symmetries.  A quick look at Table 1, however,
shows that while the chiral multiplet charges under the global
symmetries agree to a remarkable extent (recall these charged fields
do not represent local fields in the conformal field theory) there
are, in each of the theories II, III, and IV, additional fields of
$\CR$-charge one which lead to a failure of anomaly matching (except
between models II and III in which the charge spectra are identical).
Thus, the anomaly coefficients computed in the perturbative limit do
not agree, and  the four models fail this standard and basic test of
duality.

A closer inspection, however, reveals that the problem is that we
are being too simple-minded in our interpretation of the physics
described by models II, III, and IV.  The perturbative calculation of
the anomaly coefficients used above is valid when all the
gauge couplings are weak.  In model I, in which all of the factors of
the gauge group are asymptotically free, we can construct an RG
trajectory ending at the fixed point, beginning in the deep UV at weak
coupling.  In each of the other three models, the one-loop beta
functions for the various factors of the gauge groups have different
signs.  Thus, a trajectory ending at the fixed point and containing a
weakly coupled point will in general not exist.  Thus, the 't Hooft
condition that anomaly coefficients are independent of scale does not
permit us to perform perturbative calculations which will hold at the
IR fixed point.  

A closer look at Table 1 reveals that some anomaly coefficients,
namely those including at least one factor of $U(1)_\CR$, agree
between the models.  This is because the only discrepancy in the table
involves chiral multiplets of $\CR$-charge one, hence chiral fermions
that are $\CR$-neutral.  In particular, this includes the coefficient
of the $U(1)_\CR^3$ anomaly, related to the central charge of the 
conformal theory and, via the \AdS /CFT duality, to the volume of
$H$.  

Following the methods of \bh, we compare the respective 
central charges and horizon volumes in the del Pezzo and 
$\BC^3/(\BZ_3 \times \BZ_3)$ orbifold theories.  We first recall some 
facts about $\CN=1$ superconformal gauge theories in $d=4$.  In these 
theories, the trace of the stress-energy tensor receives anomalous 
contributions
\eqn\anom{
\vev{T^\mu_\mu} = -a E_4 - c I_4,}
where $E_4$ and $I_4$ are terms quadratic in the Riemann tensor,
and $a$ and $c$ (the central charge) are anomaly coefficients.  If
$R_\mu$ denotes the $\CR$-symmetry current, then the anomaly coefficients $a$
and $c$ are related to the $U(1)_\CR$-gravitational and $U(1)_\CR^3$
anomalies \afgj, 
\eqn\rc{\eqalign{
&(a - c) \sim \vev{ (\partial_\mu R^\mu) T_{\alpha \beta} T_{\gamma
\delta}} \sim \sum_{\psi} \CR(\psi) \,, \cr
&(5 a - 3 c) \sim \vev{ (\partial_\mu R^\mu) R_\alpha R_\beta} \sim
\sum_{\psi} \Big[ \CR(\psi) \Big]^3 \,. \cr}}
If we evaluate these anomaly coefficients perturbatively in each of the
four models\foot{We believe that only in Model I is there a
scale at which this perturbative analysis is valid, and it seems to be
a coincidence that the other models agree.},  summing over the $6 N^2$ 
gluinos of $\CR$-charge $1$ and two sets of $6 N^2$ matter fermions of
$\CR$-charges $-{2 \over 3}$ and $-{1 \over 3}$, we see that $a=c$ and
$c \sim 2 N^2$.

If we perform the same calculation of the central charge in the $\BC^3/(\BZ_3
\times \BZ_3)$ orbifold gauge theory, we find $c \sim 4 N^2$.  As
observed by Gubser \g, in such a situation where we deform the
orbifold gauge theory in the UV and flow to the IR \dP\ theory, we
expect a relation from the \AdS/CFT correspondence
\eqn\cv{
{{c_{IR}} \over {c_{UV}}} = {{1/\hbox{Vol}(H)} \over {1
/\hbox{Vol}(S^5/(\BZ_3 \times \BZ_3))}}}
between ratios of central charges and respective volumes of horizons.
We have just computed $c_{IR} : c_{UV}$ as $1:2$.  In section 3, we
will compute the ratio of volumes 
$\hbox{Vol}(S^5/(\BZ_3 \times \BZ_3)) : \hbox{Vol}(H)$, which we find 
to be $1:2$ as well.

\subsec{$\CN=1$ duality in the four models.}

In this section, we show that in fact the four models are related by
the $\CN=1$ duality of \sem .  This duality, of course, relates
SQCD-like theories flowing to the same conformal fixed point in the
deep IR.  Moreover, it provides a mapping of chiral operators between
the two models that can be used to map deformations of one into
``dual'' deformations of the other, preserving the property that the
physics at extremely low energies is identical.  What we mean by
$\CN=1$ duality in the context of our models is the following.
Consider, in any of the models, a factor of the gauge group under
which a total of $2N$ chiral multiplets transform in the fundamental
representation.  If we deform the theory by turning off the
superpotential couplings for these fields, as well as the gauge
coupling for any other factors in the gauge group under which they are
charged, we isolate a SQCD-like theory.  In this deformed theory,
\sem\ predicts an equivalent description in terms of ``magnetic''
variables transforming under a magnetic gauge group $SU(N)$ (hence the
insistence on $2N$ fundamentals, since we wish to remain with quivers
of this type).  One can then turn on the remaining
couplings, rewritten in terms of the magnetic variables, returning to
a dual description of the original theory.  One of the most striking
properties of the $\CN=1$ duality is the fact that such mappings of
deformations lead to agreement between two dual models.  We will see
that this is indeed the case.

So in order to apply the $\CN=1$ duality to our models with a product
gauge-group and superpotential, we start in Model I by going to a
point\foot{Note that this point can only be reached via relevant
deformations of the theory, and hence lies away from the critical
manifold of interest.  However, the dual SQCD theories will possess
dual relevant deformations which may be used to reach the critical
manifold.} corresponding to a pure $SU(N)$ SQCD theory, for which all
factors of the gauge-group but one, suppose the first, decouple and
the superpotential is turned off.  In this pure $SU(N)$ theory we
apply the usual $\CN=1$ transformation to obtain a dual description.

Thus we define composite mesonic fields ($\mu$ an arbitrary mass scale)
\eqn\mI{\eqalign{
\mu M_{25} &= X_{21} X_{15} , \cr
\mu M_{26} &= X_{21} X_{16} , \cr
\mu M_{45} &= X_{41} X_{15} , \cr
\mu M_{46} &= X_{41} X_{16} , \cr
}} 
and introduce the dual chiral fields $\tilde{X}_{12}$,
$\tilde{X}_{14}$, $\tilde{X}_{51}$, and $\tilde{X}_{61}$.  The dual
$SU(N)$ SQCD model possesses a cubic superpotential
\eqn\wcube{
\lambda \Big\{ \tr \Big[ M_{46} \tilde{X}_{61} \tilde{X}_{14} \Big] -
\tr \Big[ M_{45} \tilde{X}_{51} \tilde{X}_{14} \Big] + 
\tr \Big[ M_{25} \tilde{X}_{51} \tilde{X}_{12} \Big] - 
\tr \Big[ M_{26} \tilde{X}_{61} \tilde{X}_{12} \Big] \Big\} \,.}
The relative signs in \wcube\ are fixed by the presence in the
original SQCD theory of a global $SU(2) \times SU(2)$ symmetry under
which $X_{51}$ and $X_{61}$ (resp. $X_{12}$ and $X_{14}$) transform as
doublets.

We now restore the original superpotential and the remaining gauge
couplings. In the dual model this leads to the superpotential
\eqn\wIdual{\eqalign{
\widetilde{W}_{I} &= h_1 \tr \Big[X_{32} X_{26} X_{63}\Big] + 
h_2 \mu \, \tr\Big[M_{45} X_{54}\Big] + 
h_3 \mu \, \tr\Big[X_{34} X_{42} M_{26} X_{65} X_{53}\Big] \cr
&- h_4 \mu \, \tr\Big[M_{46} X_{63} X_{34}\Big] - 
h_5 \mu \, \tr\Big[M_{25} X_{53} X_{32}\Big] - 
h_6 \tr\Big[X_{54} X_{42} X_{26} X_{65}\Big] \cr
&+ \lambda \Big\{ \tr \Big[ M_{46} \tilde{X}_{61} \tilde{X}_{14} \Big]
- \tr \Big[ M_{45} \tilde{X}_{51} \tilde{X}_{14} \Big] + 
\tr \Big[ M_{25} \tilde{X}_{51} \tilde{X}_{12} \Big] - 
\tr \Big[ M_{26} \tilde{X}_{61} \tilde{X}_{12} \Big] \Big\} \, . \cr
}} 
We now integrate-out the massive fields $M_{45}$ and $X_{54}$ from
\wIdual\ to obtain an equivalent effective potential for the
low-energy dual theory
\eqn\lwIdual{\eqalign{
\widetilde{W}^{\hbox{\it{eff}}}_{I} &= 
h_1 \tr \Big[ X_{32} X_{26} X_{63} \Big] + 
h_3 \mu \, \tr\Big[X_{34} X_{42} M_{26} X_{65} X_{53}\Big] \cr
&- h_4 \mu \, \tr\Big[M_{46} X_{63} X_{34}\Big] - 
h_5 \mu \, \tr\Big[M_{25} X_{53} X_{32}\Big] - 2 {{h_6 \lambda} \over
{h_2 \mu}} \, \tr\Big[\tilde{X}_{51} \tilde{X}_{14} X_{42} X_{26} X_{65}\Big] \cr
&+ \lambda \tr \Big[ M_{46} \tilde{X}_{61} \tilde{X}_{14} \Big] + 
\lambda \tr \Big[ M_{25} \tilde{X}_{51} \tilde{X}_{12} \Big] - 
\lambda \tr \Big[ M_{26} \tilde{X}_{61} \tilde{X}_{12} \Big] \, . \cr
}}
We observe that $\widetilde{W}^{\hbox{\it{eff}}}_{I}$ is the
superpotential for Model II and that the representation contents of
the models agree.   Thus deforming back to the original model, we have
found that Model I is dual to Model II.  We also
observe that the fundamental chiral superfields $Y_{26}$, $Y_{46}$,
and $X_{25}$ in Model II arise under the duality
from the mesonic fields $M_{26}$, $M_{46}$, and $M_{25}$ in Model I.
Under the inverse duality transformation, we of course obtain Model I
again.  In this case we find that the fundamental field $X_{54}$ in
Model I arises from a mesonic composite field in Model II.

We now consider the relation between Model II and Model III, with
similar reasoning.  In this case, we consider the SQCD point in the
deformation space of Model II for which all $SU(N)$ factors decouple but
the one labelled '5' and for which the superpotential is turned off.
At this point, we again define the composite mesonic fields
\eqn\mII{\eqalign{
\mu \, M_{21} &= X_{25} X_{51} \, , \cr
\mu \, M_{23} &= X_{25} X_{53} \, , \cr
\mu \, M_{61} &= X_{65} X_{51} \, , \cr
\mu \, M_{63} &= X_{65} X_{53} \, , \cr
}}
and the dual fields $\tilde{X}_{15}$, $\tilde{X}_{35}$,
$\tilde{X}_{52}$, and $\tilde{X}_{56}$.  The dual superpotential
$\widetilde{W}_{II}$ then is
\eqn\wIIdual{\eqalign{
\widetilde{W}_{II} &= h_1 \tr\Big[Y_{46} Y_{61} X_{14}\Big] + 
h_2 \mu \, \tr\Big[ M_{21} X_{12}\Big] - 
h_3 \tr\Big[Y_{46} X_{63} X_{34}\Big] - 
h_4 \mu \, \tr\Big[M_{23} X_{32}\Big] \cr
&+ h_5 \tr\Big[X_{32} X_{26} X_{63}\Big]  - 
h_6 \tr\Big[X_{12} Y_{26} Y_{61}\Big] + 
h_7 \mu \, \tr\Big[X_{34} X_{42} Y_{26} M_{63} \Big] - 
h_8 \mu \, \tr\Big[X_{14} X_{42} X_{26} M_{61}\Big] \cr
&+ \lambda \Big\{ \tr \Big[ M_{23} \tilde{X}_{35} \tilde{X}_{52} \Big]
- 
\tr \Big[ M_{63} X_{35} X_{56} \Big] + 
\tr \Big[ M_{61} X_{15} X_{56} \Big]  - 
\tr \Big[ M_{21} \tilde{X}_{15} \tilde{X}_{52} \Big] \Big\} \, . \cr
}} 
Integrating-out the massive fields $M_{21}$, $X_{12}$, $M_{23}$,
and $X_{32}$ yields a low-energy effective superpotential
\eqn\lwIIdual{\eqalign{
\widetilde{W}^{\hbox{\it{eff}}}_{II} &= h_1 \tr\Big[Y_{46} Y_{61}X_{14}\Big] - 
h_3 \tr\Big[Y_{46} X_{63} X_{34}\Big] + 
\lambda \tr \Big[ M_{61} X_{15} X_{56} \Big] - 
\lambda \tr \Big[ M_{63} X_{35} X_{56} \Big]  \cr
&+ h_7 \mu \, \tr\Big[X_{34} X_{42} Y_{26} M_{63} \Big] - 
h_8 \mu \, \tr\Big[X_{14} X_{42} X_{26} M_{61}\Big] \cr 
&+ 2 {{h_5 \lambda} \over {h_4 \mu}} \, \tr 
\Big[\tilde{X}_{35} \tilde{X}_{52} X_{26} X_{63} \Big] - 
2 {{h_6 \lambda} \over {h_2 \mu}} \, \tr 
\Big[\tilde{X}_{15} \tilde{X}_{52} Y_{26} Y_{61} \Big] \, . \cr 
}}
We again see that this effective superpotential is the superpotential
of Model III and the representation contents agree.  Further, we note
that the fundamental superfields $X_{61}$ and
$Y_{63}$ of Model III arise under the duality from the composite
fields $M_{61}$ and $M_{63}$ of Model II.  Under the inverse
duality transformation, the fields $X_{32}$ and $X_{12}$ in Model II
arise as composite mesons in Model III.

Finally, to relate Model III to Model IV, we decouple all the $SU(N)$
factors of the gauge group in Model II except factor '2' and turn off
the superpotential.  At this point we define the composite mesonic
fields
\eqn\mIII{\eqalign{
\mu \, M_{46} &= X_{42} X_{26} \, , \cr
\mu \, M_{56} &= X_{52} X_{26} \, , \cr
\mu \, N_{46} &= X_{42} Y_{26} \, , \cr
\mu \, N_{56} &= X_{52} Y_{26} \, , \cr
}}
and introduce the dual fields $\tilde{X}_{24}$, $\tilde{X}_{25}$,
$\tilde{X}_{62}$, and $\tilde{Y}_{62}$.  The corresponding dual
superpotential is
\eqn\wIIIdual{\eqalign{
\widetilde{W}_{III} &= h_1 \tr\Big[ X_{56} X_{61} X_{15}\Big] - 
h_2 \tr\Big[X_{56} Y_{63} X_{35} \Big] + 
h_3 \tr\Big[ Y_{46} Y_{61} X_{14} \Big] - 
h_4 \tr\Big[ Y_{46} X_{63} X_{34} \Big] \cr
&+ h_5 \mu \, \tr\Big[ N_{46} Y_{63} X_{34} \Big] - 
h_6 \mu \, \tr\Big[ N_{56} Y_{61} X_{15} \Big] \cr
&+ h_7 \mu \, \tr\Big[ M_{56} X_{63} X_{35} \Big] - 
h_8 \mu \, \tr\Big[ M_{46}  X_{61} X_{14} \Big] \cr
&+ \lambda \Big\{ \tr\Big[ M_{46} \tilde{Y}_{62} \tilde{X}_{24} \Big]
- 
\tr\Big[ M_{56} \tilde{Y}_{62} \tilde{X}_{25} \Big] + 
\tr\Big[ N_{56} \tilde{X}_{62} \tilde{X}_{25} \Big] - 
\tr\Big[ N_{46} \tilde{X}_{62} \tilde{X}_{24} \Big] \Big\} \, . \cr 
}} 
We identify $\widetilde{W}_{III}$ with the superpotential of Model IV
and note that the representations also agree.  We also note that in
this case the fundamental fields $Y_{56}$, $Z_{56}$, $X_{46}$, and
$Z_{46}$ of Model IV have appeared from the composite fields $M_{56}$,
$N_{56}$, $M_{46}$, and $N_{46}$.  Under the inverse duality
transformation, we reproduce Model III.  None of the fundamental
fields in Model III arise as composites in Model IV.

So, by going to various points at which we can apply the $\CN=1$
duality, we see that the four models do describe the same family of
conformal theories.  In \fhhps\ the authors mention two other pairs of
``toric dual'' theories, describing D3-branes transverse to a
cone over a del Pezzo surface $dP_2$ which is the blowup of $\BP^2$ 
at two points and to a cone over a Hirzebruch surface $\BF_0$ 
(better known as $\BP^1 \times \BP^1$).  We will indicate how these 
theories are related via $\CN=1$ duality in Section 5.   We are led 
to conjecture that all instances of ``toric duality'' 
are in fact of this form.
 
\newsec{The geometry and topology of $X$ and $H$.}

In this section, we must discuss some rather detailed geometry of $X=dP_3$
and the associated non-spherical horizon $H$.  This is necessary to
understand the symmetries and especially the baryonic operator content
of the four models.  However, the result is that we will able to do a
very good job (in Section 4) of understanding the quantum ring of
baryonic operators.

We begin with the complex surface $X$, which can be described as the
blowup of $\BP^2$ at three generic (i.e. non-colinear) points.  $X$
is a smooth projective surface and a toric variety, and we have a
number of ways to understand its geometry.  We will proceed
classically, using a few facts about the geometry of non-singular
projective surfaces \gh,\hart.  (For a toric description, see for
instance \bglp, or more generally \fulton.)

\subsec{The geometry and topology of $X$.}

We start with $\BP^2$ having homogeneous coordinates $[x : y : z]$.
We blow up $\BP^2$ at three points, which without loss we may assume
to be $P_1 = [1:0:0]$, $P_2 = [0:1:0]$, and $P_3 = [0:0:1]$.  The
resulting surface is $X$, which we can present via the blow-up as a
complex subvariety of $\BP^2 \times \BP^1 \times \BP^1 \times \BP^1$.
Let $[u_1:v_1]$, $[u_2:v_2]$, and $[u_3:v_3]$ be homogeneous
coordinates on the respective $\BP^1$ factors above.  Then $X$ is the
subvariety determined by the following homogeneous equations
\eqn\blup{y v_1 - z u_1 = 0 \, , \quad z v_2 - x u_2 = 0 \, , 
\quad x v_3 - y u_3 = 0 \, .
} 
Note that away from the points $P_1$, $P_2$, and $P_3$, the three
equations in \blup\ determine a unique point in $\BP^1 \times \BP^1
\times \BP^1$, and so $X$ is locally isomorphic to $\BP^2$ on this
open set.  Conversely, for any one of those points $P_1$, $P_2$, or 
$P_3$ on $\BP^2$, one of the equations in \blup\ is identically 
satisfied for all points on one $\BP^1$ factor.  Thus on \dP, the 
points $P_1$, $P_2$, and $P_3$ of $\BP^2$ are replaced by three 
``exceptional'' $\BP^1$'s.  We also have a natural surjection, 
the blow-down, $\pi: X \rightarrow \BP^2$, which maps each of 
the three $\BP^1$'s to the points $P_1$, $P_2$, and $P_3$ and is 
the natural identification everywhere else.

We are interested really in the complex cone $C$ over $X$.  We can see
this cone explicitly if we embed $X$ as a subvariety of some
projective space $\BP^N$.  Then the homogeneous equations specifying
the image of the embedding $X \hookrightarrow \BP^N$ can also be
understood as equations for the embedding of the cone $C$ in
$\BC^{N+1}$.

To find a good embedding $C \hookrightarrow \BP^N$, we consider the
linear system of cubic curves in $\BP^2$ which pass through the three
points $P_1$, $P_2$, and $P_3$.  This system is associated to a linear
subspace $E$ of the vector space $\Gamma(\BP^2,\CO(3))$ of global
sections of $\CO(3)$.  We can easily see that a basis for $E$ is given
by the following seven sections of $\CO(3)$:
\eqn\mon{\eqalign{
&s_1=x^2 y \, , \quad s_2=x y^2 \, , \quad s_3=y^2 z \, , \cr
&s_4=y z^2 \, , \quad s_5=x z^2 \, , \quad s_6=x^2 z \, , \cr
&s_7=x y z \, . \cr}}
These seven sections determine a rational map $\BP^2 \rightarrow
\BP^6$ via $[x:y:z] \rightarrow [s_1:\ldots:s_7]$ (the map is only
rational because the sections have common zeroes at $P_1$, $P_2$, and
$P_3$).  However, we can take the divisors on $\BP^2$ corresponding to
these seven sections and then blow them up to get seven effective
divisors on $X$.  These seven divisors correspond to seven sections
(which we will also denote $s_i$) of a line-bundle on $X$, and these
seven sections on $X$ actually determine an embedding $X
\hookrightarrow \BP^6$.  The image of $X$ in $\BP^6$ corresponds to
the projective ideal $I \subset \BC[s_1,s_2,s_3,s_4,s_5,s_6,s_7]$
generated by polynomial relations arising from \mon.  This ideal has a
presentation with nine quadratic generators:
\eqn\ideal{\eqalign{
&s_1 s_7 - s_2 s_6 \, , 
\quad s_2 s_7 - s_1 s_3 \, , 
\quad s_3 s_7 - s_2 s_4 \, , \cr
&s_4 s_7 - s_3 s_5 \, , 
\quad s_5 s_7 - s_4 s_6 \, , \quad 
s_6 s_7 - s_1 s_5 \, , \cr
&s_7^2 - s_1 s_4 \; \, , 
\quad s_7^2 - s_2 s_5 \; \, , 
\quad s_7^2 - s_3 s_6. \; \, \cr
}}
So the affine coordinate ring of the cone over $X$ is $A \cong
\BC[s_1,\ldots,s_7]/I$.  In fact, we have seen the ring $A$ before;
the quadratic polynomials generating \ideal\ are precisely the ones in
\rel\ determining the chiral algebra of mesonic operators in the four
models.

We are also interested in the symmetries of $X$, since they will be
reflected in the symmetries of $C$  and the associated
horizon $H$.  By symmetries, we mean the isometries of the
K\"ahler-Einstein metric on $X$.  Although the K\"ahler-Einstein
metric on $X$ is not known explicitly, we will assume that it
respects the isometries of the K\"ahler metric on $X$ induced from
the Fubini-Study metric on $\BP^2 \times \BP^1 \times \BP^1 \times
\BP^1$ (we could equally well consider the metric arising from
$\BP^6$).  The only continuous isometries of $X$ arise from the
$U(1)^2 \times U(1) \times U(1) \times U(1)$ which acts on $\BP^2
\times \BP^1 \times \BP^1 \times \BP^1$.  Of this $U(1)^5$, only a
$U(1)^2$ subgroup respects the equations \blup\ determining $X$.  In
terms of the embedding in $\BP^6$, this $U(1)^2$ can be taken to act
with the following charges on the coordinates of $\BP^6$:
\eqn\iso{\pmatrix{
&s_1&s_2&s_3&s_4&s_5&s_6&s_7\cr
&0&-1&-1&0&1&1&0\cr
&1&1&0&-1&-1&0&0\cr
} .}

As mentioned in Section 2, $X$ also has as a discrete symmetry the
group $D_{12}$.  The $S_3$ subgroup simply arises from permutations of
the three points of the blowup on $\BP^2$.  In terms of the embedding
in $\BP^6$, it is generated by the following order-3 and order-2
permutations:
\eqn\sthree{\matrix{
&\BZ_3:&(s_1 s_3 s_5)(s_4 s_6 s_2) \, ,\cr
&\BZ_2:&(s_1 s_3)(s_4 s_6)\, . \cr}}
The other order-2 generator corresponds to the permutation
\eqn\ztwo{\matrix{
&\BZ_2:&(s_1 s_4)(s_2 s_5)(s_3 s_6)\, . \cr
}}
We will see that this $D_{12}$ is nicely realized on the set of six
lines\foot{By a line, we mean a $\BP^1$ subvariety which arises from
the intersection of hyperplanes in $\BP^6$.} on $X$.

We now discuss the divisors on $X$ and its topology in general.  $X$
has three exceptional divisors corresponding to the three $\BP^1$'s
which are blown-up, which we will denote $E_1$, $E_2$, and $E_3$.  Let
$h$ be a hyperplane in $\BP^2$, and $\pi^* h = D$ the pullback to $X$.
Then the Picard group of $X$ is freely generated by $D$, $E_1$, $E_2$
and $E_3$, so it is isomorphic to $\BZ^4$.

For this example, in fact, we have\foot{One can also see this directly
using the relative long-exact sequence for pair $(X, E_1 \cup E_2 \cup
E_3)$.} $\RH^2(X,\BZ) \cong {\rm Pic} \, X$.  The rest of the
cohomology follows from the observation that $\pi_1$ is a birational
invariant and from Poincare duality.  So
\eqn\coh{
\RH_*(X) \cong \RH^*(X, \BZ) \cong \BZ, \, 0, \, \BZ^4, \, 0, \, \BZ \, .
} 
The cup-product on $\RH^*$ is determined by the intersections of
divisors on $X$, which we may summarize as
\eqn\intx{
D \cdot D = 1 \, , \quad E_i \cdot E_j = - \delta_{ij} \, , 
\quad D \cdot E_i = 0 \, .
}

As mentioned above, $X$ has a set of six lines (these are related to
the twenty-seven lines on the cubic surface in $\BP^3$, which is also 
a del Pezzo surface).  In terms of the divisors
above, we can write the six lines as
\eqn\dv{\matrix{
&\eqalign{&L_1 = D - E_2 - E_3, \cr &L_4 = E_1, \cr}
&\eqalign{&L_2 = D - E_1 - E_3, \cr &L_5 = E_2, \cr}
&\eqalign{&L_3 = D - E_1 - E_2, \cr &L_6 = E_3. \cr}
}}
The lines $L_1$, $L_2$, and $L_3$ correspond to the blow-ups of the 
lines on $\BP^2$ passing though the points $P_2$ and $P_3$, 
$P_1$ and $P_3$, and $P_1$ and $P_2$ respectively.  The other 
lines $L_4$, $L_5$, and $L_6$ are the exceptional divisors.  In 
the homogeneous coordinates of $\BP^6$, the six lines correspond to 
the following $\BP^1$ subvarieties:
\eqn\pone{\eqalign{
&L_1=[0:0:s_3:s_4:0:0:0], \quad L_4=[s_1:0:0:0:0:s_6:0], \cr
&L_2=[0:0:0:0:s_5:s_6:0], \quad L_5=[0:s_2:s_3:0:0:0:0], \cr
&L_3=[s_1:s_2:0:0:0:0:0], \quad L_6=[0:0:0:s_4:s_5:0:0]. \cr}}
The notation above (perhaps somewhat non-standard) just means that the
lines correspond to the $\BP^1$ subvarieties of $\BP^6$ for which all
homogeneous coordinates but the indicated pairs are zero.

We see from \intx\ that each line has self-intersection $L_i^2 = -1$.
Each line also intersects transversely two other lines, as shown in
Figure 2.  The action of $D_{12}$ on the six lines is the natural one
suggested by the figure.
\midinsert
\centerline{
\epsfxsize=2.0in
\epsfbox{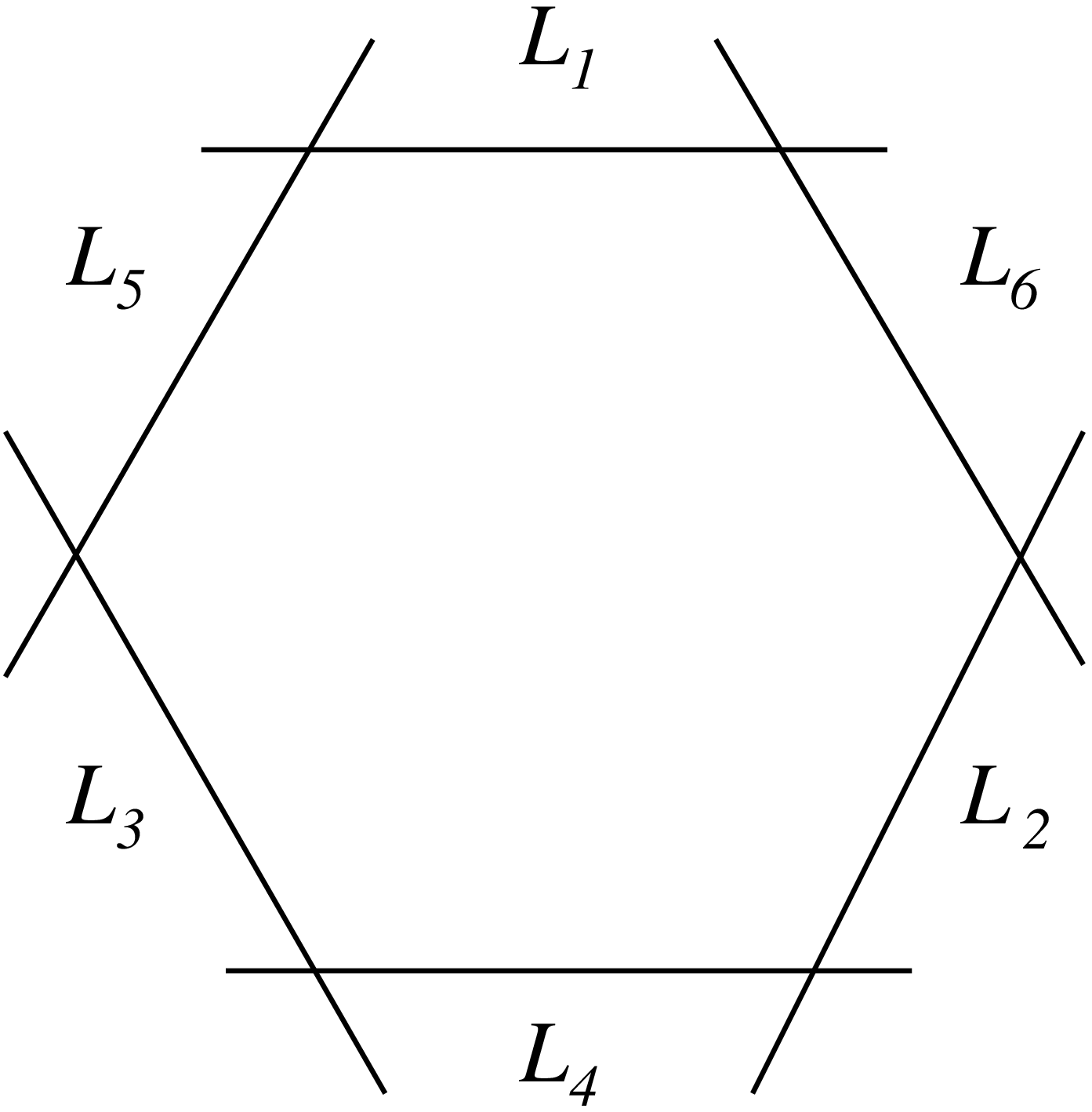}}
\centerline{Figure 2.  Six lines on \dP.}
\endinsert 

These lines also provide another way of understanding the embedding
$X \hookrightarrow \BP^6$.  We began with a linear system in
$\CO(3)$, corresponding to the divisor $3 h$, on $\BP^2$.  Away from
the points $P_1$, $P_2$, and $P_3$, $h$ is naturally isomorphic to
$\tilde{h}$, the effective divisor on $X$ which blows-down to $h$.
$\tilde{h}$ must then be the closure in $X$ of $\pi^{-1}(h - P_1 -
P_2 - P_3)$.  In terms of ${\rm Pic} \, X$ we then naturally
identify $\tilde{h} = D - E_1 - E_2 - E_3$.  Similarly, the effective
divisor on $X$ arising from the blow-up of $3 h$ is $-K = 3 D - E_1 -
E_2 - E_3$, the anti-canonical class of $X$.  So the embedding $X
\hookrightarrow \BP^6$ arises from the anti-canonical bundle $\CO(-K)$
on $X$.  In terms of the six lines, the anti-canonical class of $X$
is $-K = L_1 + \ldots + L_6$.  Let $x_1, \ldots, x_6$ be generic
global sections of the line-bundles $\CO(L_1), \ldots, \CO(L_6)$.
Since $\CO(L_1)\otimes\ldots\otimes\CO(L_6) \cong \CO(-K)$, we can
write the sections $s_i$ of $\CO(-K)$ which we used to make the
embedding in terms of the $x_i$.  We make this identification in the
obvious way.  On $\BP^2$ each section $s_i$ corresponds to a divisor
which is a linear combination of the lines through $P_2$ and $P_3$,
$P_1$ and $P_3$, and $P_1$ and $P_2$.  These divisors blow-up to the
corresponding combinations of the lines $L_1$, $L_2$, and $L_3$ on
$X$ plus the combinations of the exceptional divisors $L_4$, $L_5$,
and $L_6$ which are necessary so that the divisor is in the class of
$-K$.  For each linear combination of the lines $L_i$, we have a
corresponding product of the sections $x_i$ representing $s_i$ in
$\Gamma(X, \CO(-K))$.  Referring back to \mon, we find expressions
for the $s_i$ in terms of the ``homogeneous'' $x_i$ to be
\eqn\hom{\eqalign{
&s_1 = x_1^2 x_2 x_5 x_6^2 \, , \quad s_2 = x_1 x_2^2 x_4 x_6^2 \, ,
\quad s_3 = x_2^2 x_3 x_4^2 x_6 \, , \cr
&s_4 = x_2 x_3^2 x_4^2 x_5 \, , \quad s_5 = x_1 x_3^2 x_4 x_5^2 \, , 
\quad s_6 = x_1^2 x_3 x_5^2 x_6 \, , \cr
&s_7 = x_1 x_2 x_3 x_4 x_5 x_6 \, .\cr
}}
Note that indeed these assignments agree with the fact that $x_i$ must
vanish along the corresponding line $L_i$.  We have made this rather
extended discussion because it will be most natural to talk about the
three-cycles on $H$ which D3-branes wrap in terms of linear systems
of divisors on $X$.  We specify these divisors as the vanishing
loci of sections of line-bundles on $X$, and we write these sections
in terms of the $x_i$.  For instance, we find it
useful to record that the $U(1)^2$ isometry of $X$ may be considered
to arise from a corresponding $U(1)^2$ action on the $x_i$,
\eqn\homiso{\pmatrix{
&x_1&x_2&x_3&x_4&x_5&x_6\cr &1&0&0&0&0&-1\cr &-2&0&0&-1&1&2\cr 
}.}  
On the other hand, we have a $U(1)^4$ subgroup which acts
non-trivially on the $x_i$ but does not act on $X$ (as one sees from
\hom),
\eqn\bu{\pmatrix{
&x_1&x_2&x_3&x_4&x_5&x_6\cr
&1&-1&0&1&-1&0\cr
&0&-1&1&0&-1&1\cr
&1&1&1&-1&-1&-1\cr
&1&1&1&1&1&1\cr
}.}

\subsec{The geometry and topology of $H$.}

Of slightly more direct interest is the geometry of the horizon $H$
appearing as the azimuthal directions in the cone over $X$.  Based on
the \AdS/\CFT\ correspondence, we expect that the infrared dynamics of
the gauge theories in Section 2 are described by type IIB strings in
the background $\AdS_5 \times H$ with $N$ units of five-form flux
through $H$ \mp .  In terms of our discussion of $X$, $H$ is
described as the $S^1$-bundle in the anti-canonical bundle $\CO(-K)$
over $X$.

Since $H$ is just an $S^1$-bundle over $X$, its cohomology is
determined via the Leray spectral sequence \bt\ in terms of the
cohomology of $X$.  As discussed in \mp, the homology and cohomology
of $H$ are
\eqn\top{
\RH_*(H^5) \cong \RH^*(H^5,\BZ) \cong \BZ, 0, \BZ^3, \BZ^3, 0, \BZ \,
.
}
The element of $\RH^2(X,\BZ)$ which is sent to zero upon pullback
to $H$ is the Chern class of the $S^1$-bundle, which corresponds to
the divisor $L_1 + \ldots + L_6$.  Also, corresponding to the six
lines on $X$, $H$ has six three-cycles which wrap each $\BP^1$ on the
base and the $S^1$ in the fiber.

In terms of the dual \AdS\ theory, we will see the effect of this
topology in two places.  First, the three-cycles on $H$ lead in
the dual \AdS\ theory to $U(1)$ gauge symmetries arising from the
reduction of the self-dual R-R four-form along them.  These gauge
symmetries will appear on the worldvolume as global $U(1)$ symmetries
not corresponding to isometries of $H$.  Since $\RH_3(H) \cong
\BZ^3$, we expect a $U(1)^3$ symmetry to arise in this fashion, in
agreements with what we found in Section 2.
Second, the three-cycles of $H$ also provide for the existence of
states in the gauge theory corresponding to wrapped D3-branes.  These
D3-brane states will be charged under the $U(1)^3$ arising from the
corresponding three-cycles, and so the $U(1)^3$ should appear as a
baryonic symmetry in the gauge theories---indeed, we observed in
Section 2 this symmetry as $U(1)_C \times U(1)_D \times U(1)_E$.

The isometries of $H$ are also related to those of $X$.  (Again, we
do not know the Einstein metric on $H$, so we look at the isometries
of the metric on $H$ induced from the Euclidean metric in $\BC^7$.)
The $U(1)^2$ isometry of \iso\ lifts to $H$ with a trivial action on
the $S^1$ fibre.  Having already matched the mesonic operators in the
gauge theories to the coordinates $s_i$ of $\BC^7$, we see that indeed
the action \iso\ corresponds to the action of $U(1)_A \times U(1)_B$
on mesonic operators in \uabr.  On the other hand, $H$ has an
additional $U(1)$ isometry which arises from rotations in the $S^1$
fibre alone.  This isometry is the restriction to $H$ of the $U(1)$
action on the cone $C$ under which the coordinates $s_i$ all
have unit charge---which we recognize in the gauge theories as the
$\CR$-symmetry.  In terms of the sections $x_i$, this symmetry
corresponds to the $U(1)$ in \bu\ under which all $x_i$ have unit
charge.  The fact that this $U(1)$---which does not act on $X$---is
promoted to a symmetry of $H$ corresponds to the fact that $L_1 +
\ldots + L_6$ becomes topologically trivial on $H$ and hence should
not lead to a baryonic symmetry in the gauge theories.  The remaining
$U(1)^3$ in \bu\ which acts on the $x_i$ but not $H$ corresponds to
the baryonic symmetry in the gauge theories.

Finally, the discrete $D_{12}$ symmetry of $X$ lifts also to $H$
with a trivial action on the fibre.  The $U(1)$ isometry of $H$ 
corresponding to rotations in the fibre over $X$ is distinguished 
by the fact that it is the only subgroup of $U(1)^3$ which commutes with 
the $D_{12}$ (we similarly observe that in the gauge theory, only $U(1)_\CR$ 
in $U(1)_A \times U(1)_B \times U(1)_\CR$ commutes with the 
$D_{12}$ action in Models I and IV).  So we identify the isometry group 
of $H$ as
\eqn\isogp{ G = U(1)^3 \rtimes D_{12} \,. }
Except for the discrete symmetry, we have a direct correspondence 
between symmetries in the four gauge theories and the geometry of $H$.

We close this section by applying the method of \bh\ to compute the
volume of $H$.  The Einstein-Sasaki metric on $H$ takes the form \bg
\eqn\es{ {\bf g} = \pi^* {\bf h} + \eta \otimes \eta\, ,}
where ${\bf h}$ is the K\"ahler-Einstein metric on $X$, $\pi$ is the
projection in the $S^1$-bundle from $H$ onto $X$, and $\eta$ is a connection
1-form on the fibration with curvature $d \eta = 2 \pi^* \omega$
($\omega$ the K\"ahler form on $X$).  (More fundamentally, $\eta$ is 
the 1-form dual to the distinguished vector field on $H$ generating 
the regular $U(1)$ isometry whose orbits form the fibres of the bundle 
$H \rightarrow X$.)  Since the metric on $H$ decomposes as a direct sum
on the base $X$ and the $S^1$ fibre, it suffices to compute
separately the volume of $X$ and the length of the fibre.

The volume of $X$ is determined using the Einstein condition, which
relates the K\"ahler form to the first Chern class of $X$ (we refer 
the reader to \bh\ for a careful exposition of the ubiquitous constants 
below),
\eqn\ko{\omega = {\pi \over 3} c_1(X)\,.}  
So, using the volume form ${1 \over 2} \omega^2$, we see that
\eqn\vol{\hbox{Vol}(X) = {\pi^2 \over 18} \int_X c_1(X)^2\,.}
In terms of $\rm{H}^2(X,\BZ)$, $c_1(X) = - K = 3 D - E_1 - E_2 -
E_3$.  So the cup product \intx\ determines $\int_X c_1(X)^2 =
6$, hence $\hbox{Vol}(X) = {\pi^2 \over 3}$.

To determine the length of the fiber, we write $\eta = d \phi -
\sigma$, where $\phi$ is an angular coordinate in the fiber and $d
\sigma = 2 \omega = ({{2 \pi} \over 3}) \, c_1(X)$.  Then the fiber has
length $2 \pi$ exactly when $d \sigma = 2 \pi \, e$, where $e$ is the
Euler class (or equivalently first Chern class) of the $S^1$-bundle
over $X$.  In our case, $e = c_1(X)$, so (upon rescaling) we see 
that the length of the fiber is actually ${2 \pi} \over 3$.

Hence the volume of $H$ is $\hbox{Vol}(H) = ({{2 \pi} \over 3}) \,
\hbox{Vol}(X) = {{2 \pi^3} \over 9}$.  For the $\BC^3/(\BZ_3 \times
\BZ_3)$ orbifold, with horizon $S^5/(\BZ_3 \times \BZ_3)$, the volume
of $S^5$ in the Einstein metric is $\pi^3$, so that
$\hbox{Vol}(S^5/(\BZ_3 \times \BZ_3)) = {\pi^3 \over 9}$.  Thus,
${\hbox{Vol}(S^5/(\BZ_3 \times \BZ_3)) : \hbox{Vol}(H)} = 1:2$ 
as claimed in Section 2.

We also make the following observation, relevant for our discussion of 
wrapped D3-branes in Section 4.  Suppose that $\CL$ is a three-cycle on 
$H$ which wraps the $S^1$ in the fibre and a holomorphic curve $C$ on $X$.  
Then we can compute $\hbox{Vol}(\CL)=({{2 \pi} \over 3}) \, \hbox{Vol}(C)$ 
as above.  Since $C$ is holomorphic, $\hbox{Vol}(C) = \int_C \omega = 
\omega(C)$, which is determined by our knowledge of the class of 
$\omega$ from \ko\ and the class of $C$.  

In fact, we note that three-cycles of this form are the supersymmetric ones, 
i.e. they have minimal volume among all three-cycles in the 
same homology class.  To see this, note that any non-trivial three-cycle 
on $H$ must wrap the $S^1$ in the fibre and hence projects to some 
two-dimensional submanifold of $X$.  Because the Einstein metric 
on $H$ is a direct sum on $X$ and the fibre, it suffices to consider 
the volume of the two-manifold on $X$.  But it is well-known that 
on any K\"ahler manifold, the holomorphic submanifolds are 
volume-minimizing within their respective homology classes 
(they are calibrated by $\omega$).  Thus, for D3-branes wrapping 
supersymmetric three-cycles, we can compare the 
volumes of the wrapped three-cycles (or effectively the corresponding 
holomorphic  curves $C$ on $X$, since we are only interested in volume 
ratios) to the $\CR$-charges of the corresponding operators in the 
gauge theory.

\newsec{Baryons and wrapped D3-branes.}

In this section, we will use the $\CN=1$ duality to determine the
quantum algebra of baryonic chiral operators in the four models, at
least for operators of $\CR$-charge up to $N$.  The full chiral
algebra of operators in each model corresponds to the algebra of
functions on moduli space of the theory.  A simple-minded calculation
of this part of the chiral ring leads to results that differ among the
various models.  We will show that the chiral algebras do become
isomorphic if we include in each model quantum relations which arise
under $\CN=1$ duality from classical relations in the duals (similar
to the story in \kss).  We will further motivate the resulting
baryonic spectrum by showing that it has a natural interpretation in
terms of wrapped D3-branes in the dual \AdS\ theory (similar to
results in \refs{\w,\mp}).

\subsec{The quantum ring of baryons.}

We have already recorded in the Appendix the relevant classical
baryonic generators.  The $\CN=1$ duality, along with the requirement
that corresponding operators in dual theories must transform in the
same representations under all global symmetries, determines the
correspondence between the various baryons in the four models.  This
correspondence between baryonic operators in the four models is
indicated by corresponding labels $\CB_i$, $\CB_i(s)$, and
$\CB_i(s,t)$ in the Appendix.

A calculation shows, however, that the classically computed chiral
rings of the four models will not be isomorphic.  If we wish to claim
that they are dual, even in the weak sense of describing different
points on the same critical manifold, we will have to resolve this
puzzle.  The subtlety here \kss\ is that the classical ring may be
modified by ``quantum relations,'' extra identifications imposed by
the quantum dynamics.  In this case, as in the case studied in \kss ,
the quantum relations arise as {\it classical\/} relations in a dual
theory.  These relations arise when a composite operator 
$M_{ij} = X_{ik} X_{kj}$ in the direct model appears as a fundamental 
field $X_{ij}$ in the dual model (e.g. \mI).  Then, in the direct 
model the classical relation 
\eqn\a{M_{ij}^N = X_{ik}^N X_{kj}^N}
holds.  The dual model has dual baryonic operators 
$\tilde{X}^N_{k r}$, $\tilde{X}^N_{s k}$ corresponding to the direct 
baryons $X_{ik}^N$, $X_{kj}^N$, and similarly $X_{ij}^N$ corresponding 
to $M_{ij}^N$.  But the classical algebra of the dual model lacks 
the relation (henceforth we will be schematic, putting aside the issue
of normalizations and scales which must also enter such relations)
\eqn\b{X_{ij}^N = \tilde{X}_{k r}^N \tilde{X}_{s k}^N}
corresponding to the relation \a.  In our models, we propose such 
quantum relations to fix the operator algebras.

Since we have four models all related by chains of $\CN=1$ dualities,
if the rings of baryonic operators are to be isomorphic we must
require that all relations (quantum or classical) appearing in a given
model correspond to relations in the dual models.  Since one
generically finds new quantum relations each time one takes the dual,
one might wonder at the start whether all quantum relations can be
consistently imposed between all four models.  We now show that a set
of consistent relations can be imposed in all four models, and the
resulting algebra has a very nice geometric interpretation.

To start in Model I, there is one fundamental field $X_{54}$ which
appears as a composite field in Model II---we also see this fact in
the correspondence between baryonic operators $X_{54}^s (X_{53}
X_{34})^{N-s} \leftrightarrow (X_{51} X_{14})^s (X_{53} X_{34})^{N-s}$
under the global symmetries.  So we expect quantum relations in Model
I involving baryonic operators constructed from $X_{54}$.  Because 
Model I has a $D_{12}$ symmetry which should be preserved, we must 
symmetrize any set of quantum relations we impose.  We then 
conjecture the following quantum relations in Model I:
\eqn\qrel{\eqalign{
X_{32}^s (X_{34} X_{42})^{N-s} &= (X_{16} X_{65})^s X_{15}^{N-s} \cr
X_{26}^s (X_{21} X_{16})^{N-s} &= (X_{53} X_{34})^s X_{54}^{N-s} \cr
X_{63}^s (X_{65} X_{53})^{N-s} &= (X_{42} X_{21})^s X_{41}^{N-s} \cr
}}
and
\eqn\qreli{\eqalign{
(X_{65} X_{54})^s (X_{63} X_{34})^{t-s} (X_{65} X_{53} X_{34})^{N-t}
&= (X_{32} X_{21})^s (X_{34} X_{42} X_{21})^{t-s} (X_{34} X_{41})^{N-t} \cr
&= (X_{21} X_{16} X_{65})^s (X_{21} X_{15})^{t-s} (X_{26}
X_{65})^{N-t} \cr 
\cr
(X_{41} X_{16})^s (X_{42} X_{26})^{t-s} (X_{42} X_{21} X_{16})^{N-t}
&= (X_{53} X_{32})^s (X_{53} X_{34} X_{42})^{t-s} (X_{54} X_{42})^{N-t} \cr
&= (X_{16} X_{65} X_{53})^s (X_{15} X_{53})^{t-s} (X_{16} X_{63})^{N-t}. \cr
}}
Note that these relations are consistent with all global
symmetries.  We have found in the quantum ring six
generators of $\CR$-charge ${1 \over 3}N$,
\eqn\bi{
\CB_1, \quad \CB_2, \quad \CB_3, \quad \CB_4, 
\quad \CB_5, \quad \CB_6\,,
}
three families of generators of $\CR$-charge ${2 \over 3}N$, 
\eqn\bii{\CB_1(s), \quad \CB_2(s), \quad \CB_3(s)\,,}
and two families of generators of $\CR$-charge $N$,
\eqn\biii{\CB_1(s,t), \quad \CB_2(s,t)\,.}

Now, dualizing on the first $SU(N)$ factor implies that three
fundamental fields of Model II (i.e. $Y_{26}, X_{25}, Y_{46}$) are
composites in Model I.  So quantum relations in Model II arise
involving baryonic operators constructed from these fields.  The other
quantum relations \qrel, \qreli\ appearing in Model I must also 
have their counterparts in Model II.  So together we find 
the following relations in Model II:
\eqn\qrelii{\eqalign{
X_{32}^s (X_{34} X_{42})^{N-s} &= (X_{65} X_{51})^s Y_{61}^{N-s} \,,\cr
X_{26}^s Y_{26}^{N-s} &= (X_{53} X_{34})^s (X_{51} X_{14})^{N-s} \,,\cr
X_{63}^s (X_{65} X_{53})^{N-s} &= (X_{14} X_{42})^s X_{12}^{N-s} \,,\cr}}
and
\eqn\qrelii{\eqalign{
X_{25}^s (X_{26} X_{65})^{t-s} (Y_{26} X_{65})^{N-t} &= 
((X_{34} X_{42}) X_{14})^s (X_{34} X_{12})^{t-s} (X_{32} X_{14})^{N-t} \, , \cr
&= (Y_{61} X_{14})^s (X_{65} X_{53} X_{34})^{N-t} 
(X_{65} X_{51} X_{14})^{N-t} \,,\cr\cr
Y_{46}^s (X_{42} X_{26})^{t-s} (X_{42} Y_{26})^{N-t} &=
(X_{51} X_{12})^s (X_{53} X_{34} X_{42})^{t-s} 
(X_{51} X_{14} X_{42})^{N-t} \,, \cr
&= (X_{51} (X_{65} X_{53}))^s (Y_{61} X_{53})^{t-s} 
(X_{51} X_{63})^{N-t} \, .\cr}}
We see the same families of generators as in \bi, \bii, and \biii.

Now if we consider Model II as dual to Model III, we have two
fundamental fields in Model II which appear as composites in Model
III, namely $X_{32}$ and $X_{12}$.  However, the relations implied
from Model III already appear above as coming from Model I.  On the
other hand, in Model III there are two fundamental fields $X_{61}$ and
$Y_{63}$ which appear as composites in Model II, from which more
quantum relations arise.  To these relations, we must also add the
other relations which appeared in Model II.  So we find as quantum
relations in Model III
\eqn\qreliii{\eqalign{
X_{61}^s Y_{61}^{N-s} &= (X_{35} X_{52})^s (X_{34} X_{42})^{N-s} \,, \cr
X_{26}^s Y_{26}^{N-s} &= (X_{15} X_{34})^s (X_{35} X_{14})^{N-s} \,, \cr
X_{63}^s Y_{63}^{N-s} &= (X_{14} X_{42})^s (X_{15} X_{52})^{N-s} \,, \cr}}
and
\eqn\qreliii{\eqalign{
X_{56}^s (X_{52} X_{26})^{t-s} (X_{52} Y_{26})^{N-t} &= 
((X_{34} X_{42}) X_{14})^s ((X_{15} X_{52}) X_{34})^{t-s} 
((X_{35} X_{52}) X_{14})^{N-t} \,, \cr
&= (Y_{61} X_{14})^s (Y_{63} X_{34})^{t-s} (X_{61} X_{14})^{N-t} \,,\cr\cr
Y_{46}^s (X_{42} X_{26})^{t-s} (X_{42} Y_{26})^{N-t} &= 
(X_{15} (X_{35} X_{52}))^s (X_{15} (X_{34} X_{42}))^{t-s} 
(X_{35} (X_{14} X_{42}))^{N-t} \,, \cr
&= (X_{61} X_{15})^s (Y_{61} X_{15})^{t-s} (X_{63} X_{35})^{N-t}\,
. \cr
}}
These relations prune the baryonic operator algebra to the same
families of generators as in Models I and II.

Finally, note that none of the fields of Model III correspond to
composites of Model IV (all those fields were massive and integrated
out after dualizing), so there are no new quantum relations in Model
III arising from classical relations in Model IV.  In
Model IV there are four fundamental fields appearing as composites in
Model III (i.e. $X_{46}$, $Z_{46}$, $Y_{56}$, and $Z_{56}$).  So
besides the quantum relations which carry over from Model III, we
predict new quantum relations in Model IV involving these fields.  We
conjecture the relations
\eqn\qreliv{\eqalign{
X_{61}^s Y_{61}^{t-s} &= (X_{35} X_{24})^s (X_{34} X_{25})^{t-s} \,, \cr
X_{62}^s Y_{62}^{t-s} &= (X_{15} X_{34})^s (X_{35} X_{14})^{t-s} \,, \cr 
X_{63}^s Y_{63}^{t-s} &= (X_{14} X_{25})^s (X_{15} X_{24})^{t-s} \,, \cr}}
and 
\eqn\qreliv{\eqalign{
X_{46}^s Y_{46}^{t-s} Z_{46}^{N-t} &= 
(X_{15} X_{34} X_{25})^s (X_{15} X_{35} X_{24})^{t-s} 
(X_{35} X_{14} X_{25})^{N-t}\,,\cr
&= (X_{62} X_{25})^s (X_{61} X_{15})^{t-s} (Y_{62} X_{25})^{N-t}\,, \cr\cr
X_{56}^s Y_{56}^{t-s} Z_{56}^{N-t} &= (X_{34} X_{25} X_{14})^s 
(X_{34} X_{15} X_{24})^{t-s} (X_{35} X_{24} X_{14})^{N-t} \,, \cr
&= (Y_{61} X_{14})^s (X_{62} X_{24})^{t-s} (Y_{62} X_{24})^{N-t}\,.\cr}}
So again we see the same families of generators, and we have
consistently and simultaneously imposed in all four models all quantum
relations we expect to arise under the $\CN=1$ dualities.

\subsec{Wrapped D3-branes on $H$.}

We now explain the appearance of the six baryonic generators of
$\CR$-charge ${1 \over 3}N$, the three families of baryonic generators
of $\CR$-charge ${2 \over 3}N$, and the two families of $\CR$-charge
$N$ in terms of wrapped D3-branes on $H$.  Recall that the
$\CR$-charge of the operator in the gauge theory determines its
conformal weight at the conformal point, and hence the mass of the
corresponding state in the \AdS\ theory.  So we know that the ratios
of the masses of the corresponding brane states must be $1:2:3$.  The
mass of the wrapped brane will be proportional to the volume of the
wrapped cycle.  So we must look for families of supersymmetric three-cycles 
having volume ratio $1:2:3$ on $H$.  Equally important, we know the action 
of the global symmetries $U(1)^5 \rtimes D_{12}$ on the baryonic generators,
and, since we know how to interpret these symmetries in the \AdS\
theory, we should find the same action on the wrapped-brane states.

Consider first the six baryonic generators $\CB_1$, $\CB_2$, $\CB_3$,
$\CB_4$, $\CB_5$, and $\CB_6$ of $\CR$-charge ${1 \over 3}N$ (for
which we can take representatives in Model I $X_{21}^N$, $X_{65}^N$,
$X_{34}^N$, $X_{53}^N$, $X_{42}^N$, and $X_{16}^N$).  The $D_{12}$
symmetry acts on these operators as it acts on the six lines $L_i$ on
$X$, and hence it acts on these operators as it acts on the six
three-cycles in $H$ wrapping the lines and the fibre.  We hypothesize
then that these baryons correspond to D3-branes wrapping these
three-cycles.  As we indicated in Section 3, the supersymmetric
three-cycles on $H$ will arise in this way, wrapping a divisor on $X$ 
and the $S^1$ fibre.  To simplify notation, we will refer both to 
the divisors on $X$ and the three-cycle on $H$ by means of the sections $x_i$
(corresponding to the lines $L_i$) introduced in Section 3.  So we
propose the following correspondence of baryons and D3-branes wrapping
the following three-cycles:
\eqn\rone{\eqalign{
&\CB_1 \sim x_1 \, , \quad \CB_4 \sim x_4 \, , \cr
&\CB_2 \sim x_2 \, , \quad \CB_5 \sim x_5 \, , \cr
&\CB_3 \sim x_3 \, , \quad \CB_6 \sim x_6 \, . \cr}}
Since the lines $L_i$ are all degree 1 curves on $X$, the $S^1$ in $H$
fibres over them via the usual Hopf fibration.  So the corresponding
three-cycles on $H$ are topologically $S^3$'s.  The volumes of these 
three-cycles on $H$ are determined by the volumes of the curves 
$L_i$ on $X$, as discussed in Section 3.   We see that the volumes 
of these curves are $\hbox{Vol}(L_i) = \omega(L_i) = {\pi \over 3} \, 
(-K \cdot L_i) = {\pi \over 3}$.  (Incidentally, $-K \cdot C$ just 
determines the degree of the curve $C$ on $X$.)

We can also check that the $U(1)^5$ action on these three-cycles agrees
with the corresponding action on the baryonic states.  First, the
correspondence of baryons and wrapped branes \rone\ implies that the
$U(1)_A \times U(1)_B$ symmetry of the gauge theory corresponds to a
$U(1)^2$ action on the sections $x_i$ over $X$,
\eqn\utwo{
\pmatrix{
x_1&x_2&x_3&x_4&x_5&x_6\cr
1&0&0&0&0&-1\cr
1&0&0&1&-1&-1\cr}\, .}
Comparing to \homiso, we see that the $U(1)^2$ expected from the
geometry and the $U(1)^2$ discovered in the gauge theory are the same.

The other $U(1)_C \times U(1)_D \times U(1)_E$ subgroup is a baryonic
symmetry under which all mesons are uncharged and which does not
correspond to a geometric symmetry of $H$.  With the correspondence
\rone, the $U(1)_C \times U(1)_D \times U(1)_E$ symmetry corresponds
to a $U(1)^3$ action on the sections $x_i$ with charges
\eqn\uthree{
\pmatrix{
x_1&x_2&x_3&x_4&x_5&x_6\cr
1&-1&0&1&-1&0\cr
0&-1&1&0&-1&1\cr
1&1&1&-1&-1&-1\cr} \, .}
We see that this $U(1)^3$ baryonic symmetry is precisely the $U(1)^3$
subgroup of the $U(1)^4$ action in \bu which we predicted to appear.
Of course, the baryonic operators are homogeneously charged under the
$U(1)_\CR$ symmetry, which we have explained corresponds to the
homogeneous action of the last $U(1)$ subgroup of \bu.

We now consider baryonic operators of $\CR$-charge ${2 \over 3}N$.
There are three families of operators, $\CB_1(s)$, $\CB_2(s)$, and
$\CB_3(s)$ (for which we can consider representatives $X_{32}^s
(X_{34} X_{42})^{N-s}$, $X_{62}^s (X_{21} X_{16})^{N-s}$, and
$X_{63}^s (X_{65} X_{53})^{N-s}$ in Model I).  Under the $D_{12}$ 
symmetry, the $S_3$ subgroup permutes the three families.  
In the quantum ring, for which \qrel\ holds, the remaining $\BZ_2$ 
permutes the baryons within the three families, acting 
by $s \leftrightarrow N-s$.  So we look for three families 
of three-cycles permuted by the $S_3$ subgroup and fixed by 
the $\BZ_2$.  Based on the $\CR$-symmetry, we also look for 
families of three-cycles for which each member has twice the volume 
of the isolated three-cycles associated to the baryons of $\CR$-charge 
${1 \over 3}N$.  Finally, we note that in each family of operators 
$\CB_i(s)$ for $0 \le s \le N$, the operators $\CB_i(s=0)$ and 
$\CB_i(s=N)$ are distinguished by the fact that they are not 
really generators in the ring but occur as products of $\CR$-charge 
${1 \over 3}N$ baryons.  E.g. in Model I, 
$\CB_1(s=0) = X_{34}^N X_{42}^N = \CB_3 \CB_5$, and similarly 
for $\CB_1(s=N) = X_{32}^N = X_{16}^N X_{65}^N = \CB_6 \CB_2$ 
(note that the first relation for $\CB_1(s=0)$ is classical, 
but the relation for $\CB_1(s=N)$ is a quantum relation in Model I).

These observations lead us to make the following hypothesis 
for the correspondence of these baryonic operators to D3-branes 
wrapping moduli of three-cycles on $H$.  First, note that the 
six lines on $X$ are of course not all independent in homology, 
and we have relations in homology between pairs of lines on $X$
\eqn\pair{\eqalign{
&L_2 + L_6 = L_3 + L_5 = D - E_1\,, \cr 
&L_3 + L_4 = L_1 + L_6 = D - E_2\,, \cr 
&L_1 + L_5 = L_2 + L_4 = D - E_3\,,}}
which imply corresponding relations between three-cycles on $H$.  
So we can consider three linear systems of curves on $X$ in the 
same divisor classes as the sums of lines in \pair.  In terms of 
the sections $x_i$ on $X$, these linear systems correspond to 
the following linear combinations of global sections of 
$\CO(D-E_1)$, $\CO(D-E_2)$, and $\CO(D-E_3)$:
\eqn\pairs{
a\, x_2 x_6 + b \,x_3 x_5\,, 
\qquad a \,x_3 x_4 + b \,x_1 x_6\,, 
\qquad a \,x_1 x_5 + b \,x_2 x_4\,.}
For each such system of curves on $X$, there is a corresponding 
family of three-cycles on $H$ which wrap these curves on $X$
and the $S^1$ in the fibre.  The volumes of these three-cycles are 
determined by the volumes of the curves $(D-E_i)$ on $X$, which 
we compute as 
$\hbox{Vol}(D-E_i)={\pi \over 3}(-K \cdot (D-E_i)) = {{2 \pi} \over 3}$, 
twice the volume of the isolated curves for baryons of $\CR$-charge 
${1 \over 3}N$.

We then conjecture that the correspondence between the three 
families of baryonic operators and the corresponding 
three-cycles over these divisors is
\eqn\rtwo{\eqalign{
&\CB_1(s) \sim a \,x_2 x_6 + b \,x_3 x_5\,, \cr
&\CB_2(s) \sim a \,x_3 x_4 + b \,x_1 x_6\,, \cr
&\CB_3(s) \sim a \,x_1 x_5 + b \,x_2 x_4\,.}}
Here, the parameters $a$ and $b$ are coefficients 
specifying a point in a $\BP^1$ moduli space of 
three-cycles.  We see immediately that the $S_3$ symmetry 
of $H$ permutes these three families of three-cycles, and 
the $\BZ_2$ acts on the $\BP^1$ moduli by $[a:b] \mapsto [b:a]$.  
So the $D_{12}$ action on these three-cycles corresponds 
to the same action on the three families of baryons.

In order to discuss the $U(1)^6$ action on the wrapped-brane 
states, we will first discuss the $\BP^1$ moduli in more 
detail.  At the poles $[1:0]$ and $[0:1]$ of $\BP^1$, the 
corresponding curves in the linear systems in \pairs\ are 
reducible, and the branes wrap a pair of the individual 
$S^3$'s corresponding to baryons of $\CR$-charge ${1 \over 3}N$.  
On $X$ the corresponding divisors intersect transversely, 
so the $S^3$'s intersect along the $S^1$ in the fiber.  
The reducibility of these curves on $X$ corresponds to 
the factorization of the distinguished baryonic 
operators $\CB_i(s=0)$ and $\CB_i(s=N)$.  

At generic points in the $\BP^1$ moduli space, the 
corresponding three-cycle is topologically $\BR \BP^3$.  
One can see this by noting that the zero-locus of a 
section $a \,x_2 x_6 + b \,x_3 x_5$ for generic $a$, $b$, 
is an irreducible curve of genus zero (this follows 
from the adjunction formula), hence topologically $S^2$.  
However, this curve has degree two in $\BP^6$, which 
means that the $S^1$ fibration has twice the Chern 
class of the usual Hopf fibration---implying that 
the three-cycle on $H$ is topologically $\BR \BP^3$.

The point of this discussion is that the moduli space of the 
three-cycles about which these D3-branes wrap is not 
merely $\BP^1$, but $\BP^1$ with two distinguished points 
(the poles).  Although one might have expected the full $SU(2)$ 
symmetry of $\BP^1$ to act on the moduli, because of the 
distinguished points only a $U(1)$ acts.  Now, as explained 
in \w, the dynamics of the wrapped D3-branes will be described 
by quantum mechanics on the moduli space of wrapped three-cycles.  
Because the D3-branes are electrically charged with respect 
to the $N$ units of 5-flux, the D3-brane wavefunction will 
not be a function on $\BP^1$ but a global section 
of the line-bundle $\CO(N)$, and this bundle has $N+1$ 
sections transforming with charges $0, \ldots, N$ under 
the $U(1)$ symmetry.

This observation explains the degeneracies within the three families
of baryons.  First, the action of the $U(1)_C \times U(1)_D \times
U(1)_E$ symmetry on the baryonic operators in the gauge theory is
consistent with the associated $U(1)^3$ action induced from \uthree\
on the divisors in \rtwo.  The $U(1)_A \times U(1)_B$ geometric
symmetry is more interesting.  In each family $\CB_i(s)$ of baryons,
under one $U(1)$ generator of this symmetry all baryons are charged
equally, and under the other $U(1)$ generator the baryons within a
given family have charges $0, 1, \ldots, N-1, N$.  Comparing the
symmetries $U(1)_A \times U(1)_B$ in Table 1 to the corresponding
$U(1)^2$ action on $H$ in \utwo, we see that the $U(1) \subset U(1)_A
\times U(1)_B$ for which all baryons of a family are homogeneously
charged, acts on $H$ so that all three-cycles within a given family
transform homogeneously (which is to say, it does not act on the
$\BP^1$ moduli).  In contrast, the $U(1) \subset U(1)_A \times U(1)_B$
for which the baryons in a given family have charges $0, \ldots, N$
corresponds to a $U(1)$ action on $H$ which induces the $U(1)$ action
on the $\BP^1$ moduli.  So the baryon charges of $0, \ldots, N$ under
this symmetry are in agreement with the charges of the $N+1$ sections
of $\CO(N)$ under the $U(1)$ symmetry of $\BP^1$.

Turning to the baryons of $\CR$-charge $N$, the story is very similar.  
In Model I we take representatives of the two families of baryons 
$\CB_1(s,t)$ and $\CB_2(s,t)$ to be 
$(X_{65} X_{54})^s (X_{63} X_{34})^{t-s} (X_{65} X_{53} X_{34})^{N-t}$ 
and $(X_{41} X_{16})^s (X_{42} X_{26})^{t-s} (X_{42} X_{21}
X_{16})^{N-t}$ respectively.  In this case the $S_3$ subgroup of 
the $D_{12}$ symmetry of Model I permutes baryons within the two 
families, and the $\BZ_2$ subgroup exchanges the two families.  
Further, we note that the baryons in each family for which 
$s=t=0$, $s=N-t=0$, or $s=t=N$ factor into the product of three 
$\CR$-charge ${1 \over 3}N$ baryons.  Similarly, the baryons 
for which $s=0$, $t-s=0$, or $t=N$ factorize into the product 
of a $\CR$-charge ${1 \over 3}N$ baryon and a $\CR$-charge 
${2 \over 3}N$ baryon.

Based on the discrete and global symmetries, we propose that 
the baryon-brane correspondence is given by the following 
linear combinations of global sections of 
$\CO(2D - E_1 - E_2 - E_3)$ and $\CO(D)$:
\eqn\rthree{\eqalign{
&\CB_1(s,t) \sim ( a \,x_1 x_3 x_5 + b \,x_2 x_3 x_4 + c \,x_1 x_2 x_6) \, , \cr
&\CB_2(s,t) \sim ( a \,x_3 x_4 x_5 + b \,x_2 x_4 x_6 + c \,x_1 x_5 x_6) \, .}}
Here $[a:b:c]$ are coordinates on a $\BP^2$ moduli 
space for each family of three-cycles.  We compute the volumes of the 
curves $(2 D - E_1 - E_2 - E_3)$ and $D$ on $X$ to be 
$\hbox{Vol}(2 D - E_1 - E_2 - E_3) = \hbox{Vol}(D) = \pi$, so 
these have the proper volume to represent baryons of $\CR$-charge $N$.

We easily see that the action of the $D_{12}$ and $U(1)_C \times
U(1)_D \times U(1)_E$ symmetries in the $\CN=1$ theory is consistent
with the assignment \rthree.  To understand the action of $U(1)_A
\times U(1)_B$ on these baryons, we study more carefully the structure
of the moduli space of three-cycles above.  For instance, at the
points $[1:0:0]$, $[0:1:0]$, and $[0:0:1]$ on the $\BP^2$, we see from
Figure 2 that the representative three-cycle is a chain of three
volume-1 $S^3$'s intersecting in the $S^1$ fibers over two points on
$X$.  We see that the factorization of these
curves corresponds to the factorization of baryonic operators in the
chiral algebra.  We also see that along the three $\BP^1$'s bounding
the $\BP^2$ moduli, i.e. $[a:b:0]$, $[a:0:c]$, and $[0:b:c]$, the
representative three-cycle is one of the individual, volume-1 $S^3$'s
intersecting an $\BR \BP^3$ from the volume-2 family along the $S^1$
in the fiber.  Again, the baryonic operators have a corresponding
factorization in the quantum chiral algebra.  Finally, over a generic
point in the $\BP^2$ moduli the representative three-cycle is
topologically $S^3/\BZ_3$, the $\BZ_3$ acting homogeneously on the
embedding cooridinates $S^3 \hookrightarrow \BC^2$.  (This fact
follows because the corresponding curves on $X$ are again genus zero
but are of degree three.)  Because of the distinguished points in the
$\BP^2$ moduli space, the symmetries that act on the moduli space are
reduced from $PGL(3,\BC)$ to only $U(1) \times U(1)$.  As before, the
presence of $N$ units of 5-flux implies that the D3-brane
wavefunctions on this moduli space are sections of $\CO(N)$ on
$\BP^2$.  The global sections of $\CO(N)$ on $\BP^2$ are in 1-1
correspondence with the baryons in each family, and the geometric
$U(1)_A \times U(1)_B$ symmetry of the gauge theory acts in accordance
with the induced action of the $U(1)^2$ symmetry in \utwo\ on the
$\BP^2$ moduli.

\newsec{Other toric dualities.}

We conclude by discussing the other examples of toric duality
mentioned in
\fhhps.  These examples also arise from the $\BC^3/(\BZ_3 \times
\BZ_3)$ orbifold and correspond to cones over a del Pezzo
surface $dP_2$ and a Hirzebruch surface $\BF_0$.  $dP_2$ is
a complex surface obtained by blowing-up $\BP^2$ at two generic
points---hence it can also be obtained from \dP\ by blowing-down one
of the exceptional divisors.  As remarked earlier, $\BF_0$ is isomorphic 
to $\BP^1 \times \BP^1$.  In each of these cases, upon resolving the
orbifold we find that two distinct gauge theories arise to describe
the geometric phases corresponding to the complex cone over $dP_2$ and
the cone over $\BF_0$.

Since the cone over $dP_2$ is a resolution of the cone over \dP, there
should exist in the four \dP\ models relevant deformations which
realize the corresponding $dP_2$ models.  These deformations of course
arise from Fayet-Iliopoulos terms in the \dP\ models.  As usual,
non-zero Fayet-Iliopoulos terms in the \dP\ models imply that some
chiral fields get non-zero vevs and partially break the gauge
symmetry.  In order to see how to realize the $dP_2$ models from the
\dP\ models (other than bothering with the sort of toric algorithms in
\bglp), we just need to know which chiral fields in the \dP\ models
should get non-zero vevs.

Based on our baryon/brane correspondence, we can very easily see which
chiral fields in the \dP\ models must get non-zero vevs to realize the
$dP_2$ models.  Note that, in terms of the gauge-invariant chiral
operators, we only want to turn on vevs for baryonic operators, since
giving non-zero vevs to mesonic operators will move the D3-branes away
from the singularity.  Now, when a baryonic operator in the \dP\ gauge
theory gets a vev, a vanishing two-cycle in the cone over \dP\ is
blown-up to non-zero volume\foot{In the D3-brane theory at the
singularity, baryonic operators $\CO$ on the D3-brane worldvolume
couple to the boundaries of corresponding wrapped D3-branes via a term
$\int ds \, \CO \, \cdot \int_{C_i} \! \sqrt{g}$, where $C_i$ is a
(possibly vanishing) three-cycle.  So a vev for $\CO$ acts as a source 
for $\hbox{Vol}(C_i)$.}.  In fact, to resolve the cone over \dP\ to the cone
over $dP_2$, we want to blow-up any one of six lines on \dP\ (the
$D_{12}$ symmetry implies we needn't care which line).  D3-branes
wrapping the corresponding three-cycles on $H^5$ appear in the
gauge-theories as the six baryonic operators of $\CR$-charge ${1 \over
3}N$.  So we see that in order to realize the $dP_2$ models from the
\dP\ models, we just need to turn on a vev for any one of the six
chiral fields of $\CR$-charge $1 \over 3$.

So we first expect that the $dP_2$ models should have gauge group
$SU(N)^5$ (since only one vev is turned on), which also follows
directly from the toric algorithm.  Let us record here the
superpotentials (also appearing in \fhhps) of the two $dP_2$ models
which arise from the $\BC^3/(\BZ_3 \times \BZ_3)$ orbifold (we use the
same notation for bifundamental chiral matter as in the \dP\ models,
so we are also implicitly describing the representation content of the
$dP_2$ models)
\eqn\dptwo{\eqalign{
W_{I} &= h_1\,\tr \Big[X_{23} X_{35} X_{52}\Big] - h_2 \,\tr
\Big[X_{23} Y_{31} X_{12}\Big] - h_3 \,\tr\Big[Y_{43} X_{35}
X_{54}\Big] \cr &+ h_4 \,\tr \Big[X_{43} Y_{31} X_{15} X_{54}\Big] +
h_5 \,\tr \Big[Y_{43} X_{31} X_{12} X_{24}\Big] \cr &- h_6 \,\tr
\Big[X_{52} X_{24} X_{43} X_{31} X_{15} \Big]\,, \cr\cr W_{II} &= h_1
\,\tr \Big[ Y_{41} X_{15} X_{54} \Big] - h_2 \,\tr \Big[ Y_{41} Y_{13}
X_{34} \Big] + h_3 \,\tr \Big[ X_{21} Y_{15} X_{52} \Big] - h_4 \,\tr
\Big[ X_{21} X_{13} X_{32} \Big] \cr &+ h_5 \,\tr \Big[ Z_{41} X_{13}
X_{34} \Big] - h_6 \,\tr \Big[ X_{41} Y_{15} X_{54} \Big] \cr &+ h_7
\,\tr \Big[ X_{41} Y_{13} X_{32} X_{24} \Big] - h_8 \,\tr \Big[ Z_{41}
X_{15} X_{52} X_{24} \Big]\,.}}

Now, in \dP\ Model I, if we turn on a vev for any chiral superfield of
$\CR$-charge $1 \over 3$, it is easy to see that the resulting
low-energy effective theory describing the cone over $dP_2$ is
the first $dP_2$ model.  Similarly, in \dP\ Model IV, turning on a vev 
for any of the chiral fields of $\CR$-charge $1 \over 3$ results 
in the second $dP_2$ model.  These blow-ups from $\CN=1$
dual \dP\ models imply that the $dP_2$ models are also related by
$\CN=1$ duality.  One can also see this directly by applying the
$\CN=1$ duality to $SU(N)$ factor '1' or '4' in the first $dP_2$
model.

We also observe that in the \dP\ Models II and III (for which the
discrete $D_{12}$ is broken to $\BZ_2 \times \BZ_2$, and for which
there are now two orbits of baryons of $\CR$-charge ${1 \over 3}N$),
we can turn on vevs of $\CR$-charge ${1 \over 3}N$ operators to
realize either of the two $dP_2$ models from each \dP\ model.  In \dP\
Model II, vevs for the baryons $X_{14}^N$, $X_{34}^N$, $X_{53}^N$, or
$X_{51}^N$ yield the first $dP_2$ model, while vevs for $X_{42}^N$ and
$X_{65}^N$ yield the second $dP_2$ model.  In \dP\ Model III, vevs for
the baryons also $X_{14}^N$, $X_{34}^N$, $X_{15}^N$, or $X_{35}^N$
also yield the first $dP_2$ model, while vevs for the baryons
$X_{52}^N$ or $X_{42}^N$ yield the second $dP_2$ model.

We can also realize the pair of $\BF_0$ models from the \dP\ models.
In this case, the cone over $\BF_0$ is obtained from the cone over
\dP\ by blowing-up any of the three pairs of lines $(L_1,L_4)$,
$(L_2,L_5)$, or $(L_3,L_6)$.  So the baryon/brane correspondence in
the \dP\ models implies that turning on vevs for the corresponding
pairs of $\CR$-charge ${1 \over 3}N$ baryons will realize these
$\BF_0$ models.  For completeness, we record here the two
superpotentials of the $\BF_0$ models (the first of which appears in
\mp, both of which appear in \fhhps):
\eqn\fzero{\eqalign{
W_{I} &= h_1 \, \tr \Big[ X_{14} X_{42} Y_{23} Y_{31} \Big] + h_2 \,
\tr \Big[ Y_{14} Y_{42} X_{23} X_{31} \Big] \cr &- h_3 \, \tr \Big[
Y_{14} X_{42} X_{23} Y_{31} \Big] - \tr \Big[ X_{14} Y_{42} Y_{23}
X_{31} \Big] \,, \cr\cr W_{II} &= h_1 \, \tr \Big[ x_{21} y_{14}
y_{42} \Big] + h_2 \, \tr \Big[ y_{21} y_{13} x_{32} \Big] + h_3 \,
\tr \Big[ w_{21} x_{13} y_{32} \Big] + h_4 \, \tr \Big[ z_{21} x_{14}
x_{42} \Big] \cr &- h_5 \, \tr \Big[ w_{21} y_{14} x_{42} \Big] - h_6
\, \tr \Big[ y_{21} x_{14} y_{42} \Big] - h_7 \, \tr \Big[ x_{21}
y_{13} y_{32} \Big] - h_8 \, \tr \Big[ z_{21} x_{13} x_{32} \Big] \,
.}}

We can easily check that the first $\BF_0$ model arises from the \dP\
Model I, and the second $\BF_0$ model arises from \dP\ Model IV.  (Of
course, we could also consider the $\BF_0$ models as arising from
deformations of the corresponding $dP_2$ models as well.)  So again
the two $\BF_0$ models are related by $\CN=1$ duality.  In fact,
applying a $\CN=1$ duality transformation to any $SU(N)$ factor of the
gauge group of the first $\BF_0$ model results in the second $\BF_0$
model.

\newsec{Conclusion.}

The fact that all three of the known examples of toric duality really
arise from Seiberg's $\CN=1$ duality leads us to conjecture that this
result is general, for all examples of toric duality (and surely there
are more) arising from partial resolutions of abelian orbifolds.  If
such a conjecture is true, it reassures us of the validity of the
general procedures used to derive the worldvolume gauge theory
descriptions of D3-branes at singularities which arise as partial
resolutions of abelian orbifolds.

The gauge theories we encounter in our discussion are interesting
nontrivial examples of dual theories, exhibiting the features of
dangerously irrelevant operators, quantum modifications of chiral
rings, and RG trajectories that do not come near free fixed points.
It would be interesting to pursue the study of these theories further,
along the lines of \kss , establishing the precise operator mapping
(with normalizations and scales).  In particular, the structure of the
quantum corrected chiral ring deserves closer study.  There should be
a deformation of this corresponding to reducing the coupling constant
such that a limiting form reproduces the classical ring, at least in
the case of an asymptotically free theory such as Model I.

It would also be interesting to understand better the holographic
aspects of these systems.  We understand how the conformal field
theory is dual to an \AdS\ compactification.  Does the dynamics of the
gauge theory at intermediate scales (below the Higgs scale but above
the masses of lightest massive states) have a useful interpretation in
terms of the worldvolume theory or the holographic dual?  For instance, 
the duality cascade of \ks\ is an example of very interesting physics 
at these scales.  We leave such questions for further work.

\bigbreak\bigskip\bigskip\centerline{{\bf Acknowledgements}}\nobreak
The authors would like to thank Brian Greene and Calin Lazaroiu for
their collaboration in the earlier stages of this project.  We would
also like to thank David Morrison for helpful conversations.  C.E.B. 
would like to thank Aaron Bergman and Chris Herzog for discussions of their 
volume computations on Einstein-Sasaki fivefolds, and he would like to thank 
Igor Klebanov for pointing out the relevance of those computations to 
this project.  The work of C.E.B. is supported by an NSF Graduate Research 
Fellowship and under NSF grant PHY-9802484.  M.R.P. would like to 
acknowledge the hospitality of the Institute for Theoretical Physics, 
Santa Barbara, where part of this work was completed during the Duality 
Workshop, and thank the organizers of the workshop.  In particular, 
a conversation with M. Strassler was helpful in encouraging us to complete the
project.  The work of M.R.P. is supported in part by NSF grant
DMS-0074072.

\listrefs

\newsec{Appendix.}

\subsec{Chiral operators in Model I.}
The ring of mesonic operators is Model I is generated by 
traces of (powers of) the following seven adjoint chiral 
operators in Model I (the equalities below follow from 
the $F$-term equations\foot{We have assumed that the 
couplings $h_i$ appearing in $W$ are real and positive---otherwise, 
a field redefinition may be used to realize this situation.} ):
\eqn\mesoni{\eqalign{
\CM_1 &= X_{15} X_{53} X_{34} X_{41} =  X_{42} X_{26} X_{65} X_{53}
X_{34} \, , 
\quad \CM_4 = X_{16} X_{63} X_{32} X_{21} = X_{16} X_{65} X_{54} X_{42} X_{21} \, , \cr\cr
\CM_2 &= X_{65} X_{53} X_{32} X_{26} = X_{65} X_{53} X_{34} X_{41}
X_{16} \, , 
\quad \CM_5 = X_{54} X_{42} X_{21} X_{15} = X_{16} X_{63} X_{34} X_{42} X_{21} \, , \cr\cr
\CM_3 &= X_{54} X_{41} X_{16} X_{65} = X_{53} X_{32} X_{21} X_{16}
X_{65} \, , 
\quad \CM_6 = X_{34} X_{42} X_{26} X_{63} = X_{34} X_{42} X_{21} X_{15} X_{53} \, , \cr\cr
\CM_7 &= X_{32} X_{26} X_{63} = X_{41} X_{15} X_{54} = X_{41} X_{16} X_{63} X_{34} 
= X_{21} X_{15} X_{53} X_{32} = X_{54} X_{42} X_{26} X_{65} \, , \cr
&= X_{34} X_{42} X_{21} X_{16} X_{65} X_{53}\, . \cr}}
The classical baryonic generators in Model I of $\CR$-charge ${1 \over 3}N$ are
\eqn\bi{
\CB_1 = X_{21}^N \, , \quad \CB_2 = X_{65}^N \, , \quad \CB_3 =
X_{34}^N \, 
,\quad \CB_4 = X_{53}^N \, , \quad \CB_5 = X_{42}^N \, , \quad \CB_6 = X_{16}^N \, .}
The classical baryonic generators in Model I of $\CR$-charge ${2 \over 3}N$ are
\eqn\bbi{\eqalign{
& \CB_1(s) = X_{32}^s (X_{34} X_{42})^{N-s}\,, 
\quad \CB_2(s) = X_{26}^s (X_{21} X_{16})^{N-s}\,, 
\quad \CB_3(s) = X_{63}^s (X_{65} X_{53})^{N-s} \cr
& \CB'_1(s) = X_{15}^s (X_{16} X_{65})^{N-s}\,, 
\quad \CB'_2(s) = X_{54}^s (X_{53} X_{34})^{N-s}\,, 
\quad  \CB'_3(s) = X_{41}^s (X_{42} X_{21})^{N-s}\,. \cr}}
The classical baryonic generators in Model I of $\CR$-charge $N$ are
\eqn\bbbi{\eqalign{
& \CB_1(s,t) = (X_{65} X_{54})^s (X_{63} X_{34})^{t-s} (X_{65} X_{53} X_{34})^{N-t}\,, \cr
& \CB'_1(s,t) = (X_{34} X_{41})^s (X_{32} X_{21})^{t-s} (X_{34} X_{42} X_{21})^{N-t}\,, \cr
& \CB''_1(s,t) = (X_{21} X_{15})^s (X_{26} X_{65})^{t-s} (X_{21} X_{16} X_{65})^{N-t}\,, \cr
& \CB_2(s,t) = (X_{41} X_{16})^s (X_{42} X_{26})^{t-s} (X_{42} X_{21} X_{16})^{N-t}\,, \cr
& \CB'_2(s,t) = (X_{54} X_{42})^s (X_{53} X_{32})^{t-s} (X_{53} X_{34} X_{42})^{N-t}\,, \cr
& \CB''_2(s,t) = (X_{15} X_{53})^s (X_{16} X_{63})^{t-s} (X_{16} X_{65} X_{53})^{N-t}\,. \cr}}

\subsec{Chiral operators in Model II.}
Similarly for Model II, we have mesonic generators
\eqn\mesonii{\eqalign{
\CM_1 &= X_{12} X_{26} Y_{61} =  X_{65} X_{53} X_{34} X_{42} X_{26} \, , \cr\cr
\CM_2 &= X_{53} X_{32} X_{26} X_{65} = X_{51} X_{12} X_{26} X_{65} = X_{53} X_{34} Y_{46} X_{65} \, , \cr\cr
\CM_3 &= X_{53} X_{32} Y_{26} X_{65} = X_{51} X_{12} Y_{26} X_{65} = X_{51} X_{14} Y_{46} X_{65} \, , \cr\cr
\CM_4 &= X_{63} X_{32} Y_{26} = X_{65} X_{51} X_{14} X_{42} Y_{26} \, , \cr\cr
\CM_5 &= X_{34} X_{42} Y_{26} X_{63} = X_{14} X_{42} X_{25} X_{51} = X_{14} X_{42} Y_{26} Y_{61} \, , \cr\cr
\CM_6 &= X_{34} X_{42} X_{26} X_{63} = X_{14} X_{42} X_{26} Y_{61} = X_{34} X_{42} X_{25} X_{53} \, , \cr\cr
\CM_7 &= Y_{46} Y_{61} X_{14} = X_{25} X_{51} X_{12} = Y_{46} X_{63} X_{34} = X_{25} X_{53} X_{32} \, ,\cr
&= X_{32} X_{26} X_{63} = X_{12} Y_{26} Y_{61} \, , \cr
&=X_{34} X_{42} Y_{26} X_{65} X_{53} = X_{14} X_{42} X_{26} X_{65} X_{51}\, . \cr}}
The classical baryonic generators in Model II of $\CR$-charge ${1 \over 3}N$ are
\eqn\bii{
\CB_1 = X_{14}^N\, , \quad \CB_2 = X_{65}^N\,,\quad \CB_3 =
X_{34}^N\,,
\quad \CB_4 = X_{53}^N\,,\quad \CB_5 = X_{42}^N\,,\quad \CB_6 = X_{51}^N\,.}
The classical baryonic generators in Model II of $\CR$-charge ${2 \over 3}N$ are
\eqn\bbii{\eqalign{
&\CB_1(s) = X_{32}^s (X_{34} X_{42})^{N-s}\,, \quad \CB_2(s) =
X_{26}^s Y_{26}^{N-s}\,, 
\quad \CB_3(s) = X_{63}^s (X_{65} X_{53})^{N-s}\,, \cr
&\CB'_1(s) = Y_{61}^s (X_{65} X_{51})^{N-s}\,, \quad \CB'_2(s) =
(X_{51} X_{14})^s (X_{53} X_{34})^{N-s}\,, 
\quad \CB'_3(s) = X_{12}^s (X_{14} X_{42})^{N-s} \,.}}
The classical baryonic generators in Model II of $\CR$-charge $N$ are
\eqn\bbbii{\eqalign{
& \CB_1(s,t) = (X_{65} X_{51} X_{14})^s (Y_{61} X_{14})^{t-s} (X_{65} X_{53} X_{34})^{N-t} \, ,\cr
& \CB'_1(s,t) = (X_{34} X_{12})^s (X_{32} X_{14})^{t-s} ((X_{34} X_{42}) X_{14})^{N-t} \, ,\cr
& \CB''_1(s,t) = X_{25}^s (X_{26} X_{65})^{t-s} (Y_{26} X_{65})^{N-t} \, ,\cr
& \CB_2(s,t) = Y_{46}^s (X_{42} X_{26})^{t-s} (X_{42} Y_{26})^{N-t} \, ,\cr
& \CB'_2(s,t) = (X_{51} X_{14} X_{42})^s (X_{51} X_{12})^{t-s} (X_{53} X_{34} X_{42})^{N-t} \,, \cr
& \CB''_2(s,t) = (Y_{61} X_{53})^s (X_{51} X_{63})^{t-s} (X_{51} (X_{65} X_{53}))^{N-t} \,.\cr
}}

\subsec{Chiral operators in Model III.}
Similarly for Model III, we have mesonic generators
\eqn\mesoniii{\eqalign{
\CM_1 &= X_{26} Y_{61} X_{15} X_{52} = X_{26} Y_{63} X_{34} X_{42} \, , \cr\cr
\CM_2 &= X_{26} Y_{63} X_{35} X_{52} = X_{26} X_{61} X_{15} X_{52} = Y_{63} X_{34} Y_{46} \, , \cr\cr
\CM_3 &= Y_{26} X_{61} X_{15} X_{52} = Y_{26} Y_{63} X_{35} X_{52} = X_{61} X_{14} Y_{46} \, , \cr\cr
\CM_4 &= Y_{26} X_{61} X_{14} X_{42} = Y_{26} X_{63} X_{35} X_{52} \, , \cr\cr
\CM_5 &= Y_{26} X_{63} X_{34} X_{42} = Y_{26} Y_{61} X_{14} X_{42} = X_{63} X_{35} X_{56} \, , \cr\cr 
\CM_6 &= X_{26} Y_{61} X_{14} X_{42} = X_{26} X_{63} X_{34} X_{42} = Y_{61} X_{15} X_{56} \, , \cr\cr
\CM_7 &= X_{56} X_{61} X_{15} = X_{56} Y_{63} X_{35} = Y_{46} Y_{61} X_{14} = Y_{46} X_{63} X_{34} \, , \cr
&= X_{42} Y_{26} Y_{63} X_{34} = X_{52} Y_{26} Y_{61} X_{15} = X_{52} X_{26} X_{63} X_{35} = X_{42} X_{26} X_{61} X_{14}\,. \cr}}
The classical baryonic generators in Model III of $\CR$-charge ${1 \over 3}N$ are
\eqn\biii{
\CB_1 = X_{14}^N\,,\quad \CB_2 = X_{52}^N\,,\quad \CB_3 = X_{34}^N\,,
\quad \CB_4 = X_{15}^N\,,\quad \CB_4 = X_{42}^N\,, \quad \CB_6 = X_{35}^N\,.} 
The classical baryonic generators in Model III of $\CR$-charge ${2 \over 3}N$ are
\eqn\bbiii{\eqalign{
&\CB_1(s) = (X_{35} X_{52})^s (X_{34} X_{42})^{N-s}\,,\quad \CB_2(s) =
X_{26}^s Y_{26}^{N-s}\,,
\quad \CB_3(s) = X_{63}^s Y_{63}^{N-s}\,, \cr
&\CB'_1(s) = Y_{61}^s X_{61}^{N-s}\,,\quad \CB'_2(s) = (X_{35}
X_{14})^s (X_{15} X_{34})^{N-s}\,,
\quad \CB'_3(s) = (X_{15} X_{52})^s (X_{14} X_{42})^{N-s}\,.\cr}}
The classical baryonic generators in Model III of $\CR$-charge $N$ are
\eqn\bbbiii{\eqalign{
& \CB_1(s,t) = (X_{61} X_{14})^s (Y_{61} X_{14})^{t-s} (Y_{63} X_{34})^{N-t}\,, \cr
& \CB'_1(s,t) = (X_{34} (X_{15} X_{52}))^s ((X_{35} X_{52}) X_{14})^{t-s} ((X_{34} X_{42}) X_{14})^{N-t}\,, \cr
& \CB''_1(s,t) = X_{56}^s (X_{52} X_{26})^{t-s} (X_{52} Y_{26})^{N-t}\,, \cr
& \CB_2(s,t) = Y_{46}^s (X_{42} X_{26})^{t-s} (X_{42} Y_{26})^{N-t}\,, \cr
& \CB'_2(s,t) = (X_{35} (X_{14} X_{42}))^s (X_{15} (X_{35} X_{52}))^{t-s} (X_{15} (X_{34} X_{42}))^{N-t}\,, \cr
& \CB''_2(s,t) = (Y_{61} X_{15})^s (X_{63} X_{35})^{t-s} (X_{61} X_{15})^{N-t}\,. \cr}}

\subsec{Chiral operators in Model IV.}
Similarly for Model IV, we have mesonic generators
\eqn\mesoniv{\eqalign{
\CM_1 &= Y_{56} X_{62} X_{25} = Y_{56} Y_{61} X_{15} = X_{46} X_{62} X_{24} = X_{46} Y_{63} X_{34} \, , \cr\cr
\CM_2 &= Y_{56} X_{61} X_{15} = Y_{56} Y_{63} X_{35} = Y_{46} Y_{63} X_{34} = Y_{46} X_{62} X_{24} \, , \cr\cr
\CM_3 &= Z_{56} X_{61} X_{15} = Z_{56} Y_{63} X_{35} = Y_{46} X_{61} X_{14} = Y_{46} Y_{62} X_{24} \, , \cr\cr
\CM_4 &= Z_{56} X_{63} X_{35} = Z_{56} Y_{62} X_{25} = Z_{46} Y_{62} X_{24} = Z_{46} X_{61} X_{14} \, , \cr\cr
\CM_5 &= X_{56} Y_{62} X_{25} = X_{56} X_{63} X_{35} = Z_{46} X_{63} X_{34} = Z_{46} Y_{61} X_{14} \, , \cr\cr
\CM_6 &= X_{56} X_{62} X_{25} = X_{56} Y_{61} X_{15} = X_{46} Y_{61} X_{14} = X_{46} X_{63} X_{34} \, , \cr\cr
\CM_7 &= Y_{56} X_{63} X_{35} = Y_{56} Y_{62} X_{25} = Y_{46} Y_{61} X_{14} = Y_{46} X_{63} X_{34} \, , \cr
&= X_{56} X_{61} X_{15} = X_{56} Y_{63} X_{35} = X_{46} Y_{62} X_{24} = X_{46} X_{61} X_{14} \, , \cr
&=Z_{56} X_{62} X_{25} = Z_{56} Y_{61} X_{15} = Z_{46} Y_{63} X_{34} = Z_{46} X_{62} X_{24} \, . \cr}}
The classical baryonic generators in Model IV of $\CR$-charge ${1 \over 3}N$ are
\eqn\biv{
\CB_1 = X_{14}^N\,,\quad \CB_2 = X_{24}^N\,,\quad \CB_3 =
X_{34}^N\,,\quad \CB_4 = X_{15}^N\,,
\quad \CB_5 = X_{25}^N\,,\quad \CB_6 = X_{35}^N.} 
The classical baryonic generators in Model IV of $\CR$-charge ${2 \over 3}N$ are
\eqn\bbiv{\eqalign{
& \CB_1(s) = (X_{35} X_{24})^s (X_{34} X_{25})^{N-s}\,,\quad \CB_2(s)
= X_{62}^s Y_{62}^{N-s}\,,
\quad \CB_3(s) = X_{63}^s Y_{63}^{N-s}\,,\cr
& \CB'_1(s) = Y_{61}^s X_{61}^{N-s}\,,\quad \CB'_2(s) = (X_{35}
X_{14})^s (X_{15} X_{34})^{N-s}\,,
\quad \CB'_3(s) = (X_{15} X_{24})^s (X_{14} X_{25})^{N-s}\,.\cr}}
The classical baryonic generators in Model IV of $\CR$-charge $N$ are
\eqn\bbbiv{\eqalign{
& \CB_1(s,t) = (Y_{62} X_{24})^s (Y_{61} X_{14})^{t-s} (X_{62} X_{24})^{N-t} \,, \cr
& \CB'_1(s,t) = (X_{34} X_{15} X_{24})^s (X_{35} X_{24} X_{14})^{t-s} (X_{34} X_{25} X_{14})^{N-t} \,, \cr
& \CB''_1(s,t) = X_{56}^s Y_{56}^{t-s} Z_{56}^{N-t} \,,\cr
& \CB_2(s,t) = Y_{46}^s X_{46}^{t-s} Z_{46}^{N-t} \,,\cr
& \CB'_2(s,t) = (X_{35} X_{14} X_{25})^s (X_{15} X_{35} X_{24})^{t-s} (X_{15} X_{34} X_{25})^{N-t} \,,\cr
& \CB''_2(s,t) = (X_{62} X_{25})^s (Y_{62} X_{25})^{t-s} (X_{61} X_{15})^{N-t} \,. \cr
}}

\bye